\documentclass{jpp}

\usepackage{amsmath}
\usepackage{amssymb}
\usepackage{mathrsfs}
\usepackage{color}
\usepackage{xspace}
\usepackage[T1]{fontenc}
\usepackage{hyperref}
\usepackage[capitalize]{cleveref}
\usepackage{natbib}
\newcommand{\GX}{{GX}\xspace}

\usepackage[normalem]{ulem}
\usepackage{multirow}
\usepackage{multicol}
\usepackage{booktabs}

\renewcommand{\vec}[1]{\ensuremath{\mbox{$ {\bf #1} $}}} 

\newcommand{\uv}[1]{\ensuremath{\mathbf{\hat{#1}}}} 

\newcommand{\gm}[2] {
  \mathcal{J}_{#2} #1
}

\newcommand{\jj} {
  \ell
}

\newcommand{\kk} {
  m
}

\newcommand{\Gjk} {
  G_{\jj,\kk}
}

\newcommand{\Hev}[1] {
  {\rm He}_{#1}(\nvpar)
}

\newcommand{\Lv}[1] {
  {\rm L}_{#1}(\nmu B)
}

\newcommand{\pderiv}[2] {
  \frac{\partial #1}{\partial #2}
}

\newcommand{\gyavgR}[1] {
  \langle{ #1 \rangle}_{\bf R}
}
\newcommand{\gyavgr}[1] {
  \langle{ #1 \rangle}_{\bf r}
}

\newcommand{\vEM} {
  {\bf v}_{\chi}
}

\newcommand{\bhat}{
  {\bf \hat{b}}
}

\newcommand{\nvpar} {
  {v}_\parallel
}

\newcommand{\nmu} {
  \mu
}

\newcommand{\phys}[1] {
  #1
}

\crefname{equation}{Eq.}{}

\title{\GX: a GPU-native gyrokinetic turbulence code for tokamak and stellarator design}
\author{N.~R. Mandell\aff{1,6} \corresp{\email{nmandell@pppl.gov}},
W.~Dorland\aff{2,3},
I.~Abel\aff{3},
R.~Gaur\aff{3,5},
P.~Kim\aff{2,5},
M.~Martin\aff{4},
T.~Qian\aff{5}
}
\affiliation{
\aff{1} Princeton Plasma Physics Laboratory, Princeton, NJ 08543, USA
\aff{2} University of Maryland, College Park, MD 20742, USA
\aff{3} Institute for Research in Electronics and Applied Physics, University of Maryland,
College Park, MD 20742, USA
\aff{4} Thea Energy, Princeton, NJ 08542, USA
\aff{5} Princeton University, Princeton, NJ 08543, USA
\aff{6} Plasma Science and Fusion Center, Massachusetts Institute of Technology, Cambridge, MA
}

\begin{document}

\maketitle
\begin{abstract}
\GX is a code designed to solve the nonlinear gyrokinetic system for low-frequency turbulence in magnetized plasmas, particularly tokamaks and stellarators. {\color{black} In \GX, our primary motivation and target is a fast gyrokinetic solver that can be used for fusion reactor design and optimization along with wide-ranging physics exploration.} This has led to several code and algorithm design decisions, specifically chosen to prioritize time to solution. First, we have used a discretization algorithm that is pseudo-spectral in the entire phase-space, including a Laguerre-Hermite pseudo-spectral formulation of velocity space, which allows for smooth interpolation between coarse gyrofluid-like resolutions and finer conventional gyrokinetic resolutions {\color{black} and efficient evaluation of a model collision operator}. Additionally, we have built \GX to natively target graphics processors (GPUs), which are {\color{black} among} the fastest computational platforms available today. Finally, we have taken advantage of the reactor-relevant limit of small $\rho_*$ by using the radially-local flux-tube approach. In this paper we present details about the gyrokinetic system and the numerical algorithms used in \GX to solve the system. We then present several numerical benchmarks against established gyrokinetic codes in both tokamak and stellarator magnetic geometries to verify that \GX correctly simulates gyrokinetic turbulence in the small $\rho_*$ limit. Moreover, we show that the convergence properties of the Laguerre-Hermite spectral velocity formulation are quite favorable for nonlinear problems of interest. Coupled with GPU acceleration, which we also investigate with scaling studies, this enables \GX to be able to produce useful turbulence simulations in minutes on one (or a few) GPUs {\color{black} and higher fidelity results in a few hours using several GPUs}. \GX is open-source software that is ready for fusion reactor design studies.

\end{abstract}

\section{Introduction}

Tokamaks and stellarators
are fusion reactor concepts that use magnetic fields to reduce the rate of loss of particles, momentum, and energy from the reactor's toroidal confinement volume. 
In general, the confining magnetic fields are produced both by magnets that surround the toroidal confinement region, and by currents flowing through the confined plasma. To a great extent, the detailed spatial patterns of magnetic fields determine whether the operating conditions of a tokamak or stellarator permit  sustained thermonuclear fusion reactions. Consequently, a key step in the design of any tokamak or stellarator reactor is the evaluation of reactor performance as a function of magnetic geometry. 

Turbulent transport is one major factor that limits reactor performance. Turbulence in the fuel column is driven by the steep gradients of density, velocity, and temperature that exist between the hot, sometimes rapidly flowing, dense fuel core and the relatively cold, diffuse environment. Under optimal conditions, the turbulent fluctuations are small-amplitude perturbations of the steady-state plasma density, temperature, {\it etc.,} but these fluctuations nonetheless enhance losses, reduce insulation (by rapidly mixing cold plasma from the outer region with hot plasma in the reactor core), and generally limit reactor performance \citep{Abel2008}. Decades of meticulous experiments and direct observations of plasma fluctuations in many tokamaks and stellarators have established a good understanding of when and where the turbulence arises, how intense it is, its dominant wavelengths and frequencies, and so on. The evaluation of the expected performance of any proposed tokamak or stellarator reactor design therefore essentially requires some way to estimate properties of the plasma turbulence that will arise as a result of the details of the proposed magnetic field geometry. 

More than two decades of extensive validation campaigns have established that gyrokinetic theory quantitatively describes this turbulence. Many gyrokinetic simulation codes {\color{black} \citep{parker1993a, kotschenreuther1995, dorland2000, lin2000, jenko2000, jost2001,Candy2003,idomura2003,watanabe2005,wang2006,jolliet2007, lanti2019,idomura2008,peeters2009,Candy2016,Barnes2019}} have been used to study, test, and ultimately validate our understanding of how plasma turbulence depends on the magnetic geometry \citep[see \emph{e.g.}][]{Belli2008,Marinoni2009,Mynick2009,White2019}. Because there are many algorithmic trade-offs to consider when setting up a turbulence simulation code, 
no single gyrokinetic code is ``best'' for these purposes. In this paper we present \GX, {\color{black} a fast gyrokinetic solver where our primary motivations and targets are fusion reactor design, optimization, and wide-ranging physics exploration at reactor scales.
}

What does this mean in practice? It means primarily that we have prioritized time to solution (together with quantitative uncertainty estimates) over comprehensiveness and full realism, and that we aim to provide simple runtime control parameters to shift smoothly from very fast but relatively less accurate calculations such as are of most value in ``outer loops'' of a design activity, to more comprehensive, ``standard'' results that are (much) more expensive to evaluate. Moreover, instead of retrofitting an existing CPU-based code, we started from scratch, with every algorithmic design decision made to target graphics processors (GPUs), which are {\color{black} among} the fastest computational platforms available today{\color{black}, with NVIDIA or AMD GPUs providing most of the computer power for 7 of the top 10 fastest supercomputers in the world \citep{Strohmaier2022}}. Finally, since we are interested in reactors specifically, we focused on the reactor limit of small $\rho_* \equiv \rho_i/R$, where $\rho_i$ is the typical radius of gyration of an ion in the plasma, and $R$ is the major radius of the device. In the small $\rho_*$ limit, there are major simplifications to be had, corresponding physically to the opening of a wide ``gap'' between typical radial correlation lengths $\lambda_{\rm corr}\sim\rho_i$ of the turbulence and the radial gradient scale lengths $L\sim R$. In contrast, present-day devices are not in the small $\rho_*$ limit, so turbulent eddies can be large enough to sample variations in the driving gradients and other relevant plasma properties. {\color{black} Additionally, because stellarators are non-axisymmetric, they have irreducible geometric variation within a given flux surface that makes the small $\rho_*$ limit more challenging. The turbulent correlation length $\lambda_{\rm corr}$ must also be small compared to the distance that geometric differences within the flux surface change significantly. This limit exists for small enough $\rho_*$, but it is roughly more challenging by a factor of the aspect ratio $R/a$, where $a$ is the minor radius of the torus.} To account for the ``mesoscale'' phenomena that may result and to complete the many quantitative validation studies that have been undertaken, many teams chose algorithms that handle radially non-local physics (often called ``global'' codes). Codes that take advantage of the simpler, radially-local, small $\rho_*$ limit are known as ``flux-tube'' codes, and our new code \GX is in this category. Several flux-tube turbulence calculations can then be coupled together with a transport solver to obtain transport time-scale profile evolution on the device scale \citep{Barnes2009} {\color{black} or steady-state profiles \citep{Candy2009}}. This multi-scale approach is asymptotically valid in the limit of vanishing $\rho_*$ {\color{black} \citep{Abel2013}}. {\color{black} Recent work has also shown a novel approach to include radially-global effects within the flux-tube formulation \citep{Parra2015,St-Onge2022}.}

To verify that \GX correctly simulates gyrokinetic turbulence in the small $\rho_*$ limit, we present several tokamak and stellarator benchmarks of \GX with established flux-tube gyrokinetic simulation codes.
Note that while numerical {\it verification} (benchmarking) is code-specific, the extensive history of experimental {\it validation} of the gyrokinetic model is inherited by any \emph{verified} gyrokinetic code and need not be repeated. \GX is open-source scientific software\footnote{The source code is available at \url{https://bitbucket.org/gyrokinetics/gx}. Documentation is available at \url{https://gx.readthedocs.io/en/latest}.} that is therefore ready for immediate design studies. Below, we show that \GX can produce useful turbulence simulations in minutes using one (or a few) GPUs {\color{black} and higher fidelity results in a few hours using several GPUs}.

Our algorithmic approach to solving the gyrokinetic system is based in pseudo-spectral methods. There is a long history of the use of Fourier pseudo-spectral methods in turbulence simulations to discretize the configuration space, especially in flux-tube gyrokinetic codes like GS2 \citep{kotschenreuther1995,dorland2000}, (non-global) GENE \citep{jenko2001}, CGYRO \citep{Candy2016}, and stella \citep{Barnes2019}. These methods have the advantage of spectrally-accurate evaluation of derivatives. Building on this, we have developed a pseudo-spectral formulation for discretizing the velocity space in the gyrokinetic system using a Laguerre-Hermite spectral basis \citep{Mandell2018}. Projecting the {\color{black}(perturbed)} gyrokinetic distribution function onto this basis produces spectral modes that correspond to (gyro)fluid moment quantities (like density, parallel momentum, etc.), so that projection of the gyrokinetic equation corresponds to a gyrofluid system with arbitrarily many moments. In the lowest-resolution limit the model retains critical conservation laws (density, momentum, energy, and free energy) and corresponds precisely to the well-established gyrofluid models of \citet{Dorland1993,Beer1996,Snyder2001} {\color{black} when their Landau-fluid closures are employed \citep{Hammett1990,Beer1996}. At moderate velocity-space resolution, one can extend the Landau-fluid closure approach to accommodate more moments \citep{Smith1997}} or use ``hyper-collisions'' to close the moment hierarchy. At high velocity-space resolution, the model {\color{black} can retain a similar number of degrees of freedom as typically used in} conventional gyrokinetic Eulerian models. A key advantage of our approach is therefore the flexibility to interpolate between these limits, depending on the desired balance of accuracy and speed for a particular calculation.  Additionally, our Fourier-Laguerre-Hermite pseudo-spectral algorithm is a good fit for modern GPU computing: the algorithm relies heavily on fast transform methods that are well-optimized on GPUs, and the memory requirements of spectral algorithms are low enough to fit a problem onto one or a few GPUs. {\color{black} Several gyrokinetic codes have  been ported to GPUs in recent years to target modern heterogeneous computing platforms \citep{Madduri2011,DAzevedo2018,Sfiligoi2018,Belli2022,Germaschewski2021,Ohana2021}.}

{\color{black} Similar works on Laguerre-Hermite spectral methods for drift kinetics \citep{Jorge2017} and gyrokinetics  \citep{Jorge2019, Frei2020,Frei2021,Frei2022,Frei2022a,Hoffmann2023a,Hoffmann2023} have shown promising results and can be viewed as complementary to our work.  
In particular, extensive linear benchmarks and convergence studies have been performed \citep{Frei2022,Frei2023}, showing that the Laguerre-Hermite approach can be used successfully to model ion temperature gradient (ITG), trapped electron, kinetic ballooning, and microtearing modes. \citet{Jorge2019,Frei2021,Frei2022,Frei2022a} develop, implement, and study advanced gyrokinetic collision operators by leveraging the Laguerre-Hermite spectral approach. Nonlinear studies in Z-pinch \citep{Hoffmann2023} and full toroidal \citep{Hoffmann2023a} geometry have also now been performed, showing that accurate nonlinear results, including the Dimits shift \citep{Dimits2000}, can be obtained with fewer basis functions than typically necessary for convergence of linear instabilities.
Additionally, \citet{Frei2020} developed a Laguerre-Hermite projection of the full-$f$ gyrokinetic system to target the periphery of tokamaks, and \citet{Frei2024} implemented the full-$f$ system in slab geometry for modeling linear devices.
}

The remainder of this paper is organized as follows: In \cref{sec:gk} we describe the gyrokinetic model. We describe the Fourier-Laguerre-Hermite pseudo-spectral formulation of the gyrokinetic system in \cref{sec:lh}. \cref{sec:geo} describes the magnetic geometry and coordinates used for tokamaks and stellarators. The time discretization scheme is described in \cref{sec:time}. In \cref{sec:benchmarks}, several benchmarks are presented comparing \GX to established flux-tube gyrokinetic codes (GS2, GENE, and stella) for linear and nonlinear cases in tokamak and stellarator geometries. \cref{sec:convergence} details the convergence properties of the Laguerre-Hermite basis for nonlinear calculations, along with details about GPU performance and multi-GPU scaling. Finally, we present the conclusions and future work in \cref{sec:conclusion}.

\section{Gyrokinetic model equations} \label{sec:gk}
In \GX we solve the $\delta\!f$ gyrokinetic system.
The literature of gyrokinetic theory is vast; our notation and philosophy follow four particular references: 
\cite{Antonsen1980},
 \cite{Frieman1982}, \cite{Barnes2010}, 
and \cite{Abel2013}. 
The velocity space in gyrokinetic theory is two-dimensional because the particles gyrate very rapidly around their guiding center position, averaging over any field variations that are encountered in one gyration period so that we can ignore the instantaneous position of the particle around its gyro-orbit.
We choose to write the gyrokinetic equation in $(v_\parallel,\mu)$ coordinates, where $v_\parallel$ is the speed in the direction of the magnetic field and $\mu \equiv v_\perp^2/2B$ is the magnetic moment, where $v_\perp$ is the speed perpendicular to the magnetic field {\color{black} and $B$ is the local magnetic field strength}. Using $\mu$ as a coordinate takes advantage of the fact that $\mu$ is a gyrokinetic adiabatic invariant.  These coordinates are not optimal for resolving the boundary layers in velocity that occur in linear gyrokinetic theory where particle orbits are ``trapped'' in magnetic wells on one side of the boundary layer and ``passing'' (untrapped) on the other. To resolve the sharp boundary layers it is better to use energy and pitch angle as one's velocity space coordinates (as is done in the GS2 code, for example, which optimizes its velocity space grid to enhance resolution near the trapped-passing boundary), but this leads to a considerably more complicated code {\color{black} implementation}, particularly for stellarator applications{\color{black}, because of the need to carefully handle turning points in each magnetic well}. However, in a sufficiently turbulent plasma, these boundary layers will be broadened{\color{black}, reducing the need for coordinates optimized for a sharp boundary}. Thus, we have selected the coordinates that lead to simpler, more efficient code because we are particularly interested in calculating nonlinear turbulence properties (as opposed to linear stability conditions). These coordinates are also used in GENE and stella.

Following Eq.~(144) of \cite{Abel2013}, but expressing the equations in $(v_\parallel,\mu)$ velocity coordinates, neglecting strong equilibrium flows, and normalizing all quantities to non-dimensional form (see \cref{app:norm}), we have
\begin{align}
\pderiv{h_s}{t} & + \left[v_{ts} v_\parallel \bhat + \gyavgR{\vEM} +
                  \frac{\tau_s}{Z_s}{\bf v}_d \right] \cdot \nabla h_s - v_{ts} \mu \left(\bhat\cdot\nabla B\right) \pderiv{h_s}{v_\parallel}
                  \notag \\ &=
                 \frac{Z_s}{\tau_s}F_{Ms} \pderiv{\gyavgR{\chi}}{t} - \gyavgR{\vEM} \cdot \nabla\big|_E F_{Ms} + C(h_s).
\label{gk}
\end{align}
Here, we have split the total distribution function as $F_s = F_{0s} + \delta\!f_s = F_{Ms}(1-Z_s\Phi/\tau_s)+h_s$, with $(Z_s/\tau_s) F_{Ms} \Phi({\bf r},t)$ and $h_s(\vec{R},v_\parallel,\mu,t)$ the Boltzmann and non-Boltzmann parts of the $\delta\!f_s$ perturbation, respectively. 
{\color{black} The equilibrium distribution function is a Maxwellian in energy $E$ (which we take to have no flows),
\begin{equation}
F_{0s} = F_{Ms}  = 
\frac{n_s}{(2\pi v_{ts}^2)^{3/2}}e^{-E} = 
\frac{n_s}{(2\pi v_{ts}^2)^{3/2}}e^{-v_\parallel^2/2- \mu B},
\end{equation}
and the fluctuating gyroaveraged distribution function satisfies
$h_s\ll F_{Ms}$. Note that in \cref{gk}, the gradient of $F_M$ at constant energy is denoted $\nabla\big|_E F_M$. We also have the following dimensionless species
parameters: equilibrium density $n_s$, equilibrium temperature
$\tau_s$, mass $m_s$, charge $Z_s$, and thermal velocity
$v_{ts}=\sqrt{\tau_s/m_s}$.}

The distribution function describes the
probability of finding a particle of species $s$ with gyrocenter position ${\bf R}$, velocity parallel to the magnetic
field $v_\parallel$, and magnetic moment $\mu=v_\perp^2/2B$, where
$v_\perp$ is the speed in the plane perpendicular to the magnetic
field. The equilibrium magnetic field ${\bf B}$ has magnitude
$B=B({\bf r})$ and direction $\hat{\bf b} = {\bf B}/B$, and magnetic fluctuations are given by $\delta \vec{B} = \delta \vec{B}_\perp + \delta\!B_\parallel \uv{b}$, where the perpendicular and parallel components can be expressed in terms of the vector potential $\vec{A} = \vec{A}_\perp + A_\parallel \uv{b}$ via $\delta \vec{B}_\perp = {\color{black}  [\nabla \times \vec{A}]_\perp \simeq } \nabla A_\parallel \times \uv{b}$ and $\delta\! B_\parallel = \uv{b}\cdot(\nabla \times \vec{A}_\perp)$, respectively. The gyrokinetic potential $\chi(\vec{r},t)$ is composed of the electrostatic potential $\Phi({\bf r}, t)$ and the vector potential $\vec{A}({\bf r}, t)$,
\begin{equation}
    \chi({\bf r}, \vec{v}, t) = \Phi({\bf r}, t) - v_{ts} \vec{v}\cdot \vec{A}({\bf r}, t),
\end{equation}
and is a function of particle position $\bf r$, where
${\bf r}= {\bf R} + \boldsymbol{\rho}$ so that $\boldsymbol{\rho}$ is
the gyroradius vector that rotates at the gyrofrequency $\Omega$ and
points from the gyrocenter (at $\bf R$) to the particle (at $\bf
r$). In \cref{gk} gyroangle dependence is eliminated by gyroaveraging the potentials at fixed gyrocenter position $\vec{R}$, denoted by
\begin{equation}
    \gyavgR{\chi} = \gyavgR{\Phi} - v_{ts}v_\parallel\gyavgR{A_\parallel} - v_{ts}\gyavgR{\vec{v}_\perp\cdot\vec{A}_\perp},
\end{equation}
along with the corresponding gyroaveraged fluctuating velocity $\gyavgR{\vEM} = \uv{b}\times\nabla \gyavgR{\chi}$. For most applications, the plasma $\beta$ is low and the magnetic
drift velocity is ${\bf v}_d =\bhat\times(v_\parallel^2 \bhat\cdot\nabla\bhat + \mu\nabla B)/B$. GX normally uses this low-$\beta$ approximation, but the full expression for the magnetic drifts can be selected at runtime.

The collision operator is denoted by $C(h)$. Here we will follow \cite{Mandell2018} and take the Dougherty model collision operator \citep{Dougherty1964}. Expressed in Fourier space, the like-species Dougherty collision operator is given by
\begin{gather}
C_{ss}(h_s) 
= \nu_{ss} \left\{ \left[ {\partial \over
     \partial v_\parallel} \left({\partial \over \partial v_\parallel}
   + v_\parallel \right) 
 + 2 {\partial \over \partial \mu} \left( {\mu \over B}
     {\partial \over \partial \mu} + \mu \right) + k_\perp^2\rho_s^2
     \right] h_s  \right. \notag \\
  \left. \qquad + \left(  \bar{T}_s \left[
      (v_\parallel^2 - 1) + 2 (\mu B - 1) \right] J_{0 s} +
  \left[ \bar{u}_{\parallel s}  v_\parallel J_{0 s} + 
    \bar{u}_{\perp s} v_\perp J_{1 s} \right] \right) F_{Ms} \right\},
\label{eq:fullC}
\end{gather}
with the field-particle terms given by
\begin{gather}
     \bar{u}_{\parallel s} \equiv \int d^3{\bf v} \, J_{0 s}\, v_\parallel \, h_s, \\
     \bar{u}_{\perp s} \equiv \int d^3{\bf v} \, J_{1 s}\, v_\perp \, h_s, \\
     \bar{T}_s = {1 \over 3} (\bar{T}_{\parallel s} + 2 \bar{T}_{\perp s}) \equiv {1 \over 3} \int d^3{\bf v} \, \left[ (v_\parallel^2 - 1) + 2 (\mu B - 1) \right] J_{0 s} h_s.
\end{gather}
Here we have the Bessel functions $J_{0 s}=J_{0}(\alpha_s)$ and $J_{1 s}=J_{1}(\alpha_s)$ that result from the Fourier transform of gyroaverage operators, with $\alpha_s = \sqrt{2\mu B b_s}$, $b_s = k_\perp^2 \rho_s^2$, {\color{black} and $\rho_s = m_s v_{ts}/(Z_s B)$ is the normalized gyroradius for species $s$.}

The Dougherty collision operator is a good physical model of like-particle
collisions, which are important to gyrokinetic dynamics.  It captures
the physics of the collision operator presented in \cite{Abel2008},
except that our collision frequency does not have velocity
dependence, and there is no difference between the rates of pitch-angle scattering and slowing down.
It is also attractive because its Laguerre-Hermite transform is sparse.
More realistic collision operators could be included in our model in the future, such as those in \cite{Frei2021,Frei2022a}, which employs a similar Laguerre-Hermite spectral-velocity approach as ours. The Dougherty model operator can also be extended to multiple species \citep{Francisquez2022}.
    
The electromagnetic potentials are determined by the gyrokinetic equivalent of Maxwell's equations. The electrostatic potential is determined by the quasineutrality equation,
\begin{equation}
\sum_s \frac{ Z_s^2 n_s}{\tau_s} \Phi = \sum_s Z_s n_s \int d^3{\bf v}\ \gyavgr{h_s}, \label{qneuts}
\end{equation}
where note that here the gyroaverage on the right-hand side is taken at constant particle position $\vec{r}$.
The parallel magnetic vector potential is determined by  the parallel component of Amp\`ere's law,
\begin{equation}
     -\nabla_\perp^2 A_\parallel =\frac{\beta_\mathrm{ref}}{2}\sum_s Z_s n_s v_{ts}  \int d^3{\bf v}\ v_\parallel  \gyavgr{h_s}, \label{ampere}
\end{equation}
with $\beta_\mathrm{ref} = 8\pi n_\mathrm{ref} T_\mathrm{ref}/B_{\color{black}\mathrm{N}}^2$ the plasma beta of the reference species. The perpendicular component of Amp\`ere's law determines the perpendicular component of the vector potential, but is more conveniently expressed in terms of the parallel magnetic fluctuation, $\delta\! B_\parallel$:
\begin{gather}
     \nabla_\perp^2 \delta\! B_\parallel = -\frac{\beta_\mathrm{ref}}{2 {\color{black} B}}\nabla_\perp\nabla_\perp : \sum_s n_s \tau_s \int d^3\vec{v}\ \langle \vec{v}_\perp \vec{v}_\perp h_s\rangle_r, \label{dB}
\end{gather}
{\color{black} where the double dot product is defined as $A : B = \sum_{ij} A_{ij} B_{ij}$}.

Finally, note that we can define an auxiliary distribution function 
\begin{equation}
    g_s = h_s - \frac{Z_s}{\tau_s}\gyavgR{\chi}F_{Ms},
\end{equation}
which eliminates the time derivative on the right-hand side of \cref{gk}, resulting in
\begin{align}
\pderiv{g_s}{t} & + \left[v_{ts} v_\parallel \bhat + \gyavgR{\vEM} +
                  \frac{\tau_s}{Z_s}{\bf v}_d \right] \cdot \nabla h_s - v_{ts} \mu \left(\bhat\cdot\nabla B\right) \pderiv{h_s}{v_\parallel}
                  = - \gyavgR{\vEM} \cdot \nabla\big|_E F_{Ms} + C(h_s).
\label{gkg}
\end{align}
This is the form of the gyrokinetic equation that we will solve numerically. 

{\color{black}\section{Magnetic geometry and choice of coordinates} \label{sec:geo}}
The geometric structure of a tokamak's magnetic field is axisymmetric, meaning there are no geometric variations that distinguish points at different toroidal angles. Most (but not all) gyrokinetic studies have focused on the axisymmetric limit. Stellarator magnetic fields are not axisymmetric. We have selected coordinates that are convenient for either case. Our spatial coordinates are aligned with the local background magnetic field so that we are able to efficiently resolve perturbations whose parallel wavelengths $\lambda_\parallel$ are $\mathcal{O}(\rho_*^{-1})$ longer than the wavelengths $\lambda_\perp$, perpendicular to the local magnetic field; these field-line-following coordinates reduce the number of grid points required to resolve the turbulence by a factor of $\rho_*$. In the case of stellarators, the lack of axisymmetry implies that even when one can describe the turbulence successfully as radially local, it might happen that the geometric variations in the binormal direction (within the magnetic surface) are not numerically well-separated from $\lambda_\perp$, even for reactor-relevant values of $\rho_*$. We have chosen not to address this possibility; we assume that $\rho_*$ is small enough that the turbulence within each flux tube in a given flux surface is not affected by turbulence occurring in a different flux tube within the same surface. 

In this section, we will start with the general form of a divergence-free magnetic field, define key important quantities, various coordinate systems, and briefly explain how they are used to obtain the geometry-dependent coefficients in the gyrokinetic model. Please refer to Appendix~\ref{app:geometry-appendix1} for specific mathematical details.

\subsection{Axisymmetric configurations (tokamaks)}
We start from the Clebsch form~\citep{Dhaeseleer1991} for the equilibrium magnetic field,
\begin{equation}
    \phys{\mathbf{B}} = \phys{\nabla} \alpha\times \phys{\nabla} \phys{\psi} .
    \label{Clebsch}
\end{equation}
We restrict our attention to solutions whose magnetic field lines lie on closed nested toroidal surfaces, known as flux surfaces. Here, flux surfaces are labeled by $\psi$, a radial-like coordinate.  On each surface, the binormal coordinate $\alpha$ is a field-line label such that a line of constant $\alpha$ gives the path of a magnetic field line.

For tokamaks, we choose to define the coordinate $\psi$ to be the enclosed poloidal flux divided by $2\pi$,
\begin{equation}
    \psi = \frac{\Psi_\mathrm{pol}}{2\pi} = \frac{1}{(2\pi)^2}\int_V d\tau \,\vec{B}\cdot\nabla\theta,
\end{equation}
and the binormal coordinate $\alpha$ is chosen to be
\begin{equation}
    \alpha = \phi - q(\psi)\theta,
\end{equation}
where $\phi$ and $\theta$ are ``straight-field-line'' toroidal and poloidal angles, respectively, and 
\begin{equation}
    q(\psi)  = \frac{1}{(2\pi)^2}\int_{0}^{2\pi} d\phi \int_{0}^{2\pi} d \theta\, \frac{\boldsymbol{B}\cdot \boldsymbol{\nabla}\phi}{\boldsymbol{B}\cdot\boldsymbol{\nabla}\theta},
    \label{eqn:safety-factor}
\end{equation}
is the safety factor. 

\subsection{Non-axisymmetric configurations (stellarators)}
For stellarators, we find it more convenient to write the magnetic field in Clebsch form as
\begin{equation}
    \vec{B} = \nabla \psi \times \nabla \alpha,
\end{equation}
with the toroidal flux chosen as the radial-like coordinate, 
\begin{equation}
    \psi = \frac{\Psi_\mathrm{tor}}{2\pi} = \frac{1}{(2\pi)^2}\int_V d\tau\, \vec{B}\cdot\nabla\phi,
\end{equation}
and the binormal coordinate chosen to be
\begin{equation}
    \alpha = \theta - \iota(\psi) \phi.
\end{equation}
Here, $\iota(\psi) = 1/q(\psi)$ is the rotational transform, and once again $\phi$ and $\theta$ are ``straight-field-line'' toroidal and poloidal angles, respectively.

\subsection{Field-aligned coordinate system}
Since \GX is a local flux-tube code, it is advantageous to use a field-aligned coordinate system. Therefore, we introduce the field-aligned, flux-tube coordinates $(x, y, z)$ used by~\citet{Beer1995}. Here, $x = x(\psi)$ is a radial coordinate, $y= y(\alpha)$ is a binormal coordinate, and $z=z(\theta)$ is a field-line-following coordinate that parametrizes distance along the equilibrium magnetic field using the poloidal angle $\theta$. The coordinates $x$ and $y$ are normalized forms of $\psi$ and $\alpha$ such that
\begin{align}
    x &= \frac{dx}{d\psi}(\psi-\psi_0), \\
    y &= \frac{dy}{d\alpha}(\alpha-\alpha_0),
\end{align}
where $\psi_0$ and $\alpha_0$ are the values of equilibrium $\psi$ and $\alpha$ at the center of the flux tube. The coordinates $(x, y, z)$ have units of length. In the local flux-tube, $x$ and $y$ vary on the length scale of the gyroradius whereas $z$ varies on the length scale of the machine size. We require an equispaced coordinate $z$ along the field line, which is needed to utilize the Fourier spectral treatment of the parallel derivative terms in the gyrokinetic equation. The procedure for obtaining an equispaced, equal-arc $z$ coordinate is provided in Appendix~\ref{app:geometry-appendix1}. 

Upon defining the field-aligned coordinate system, we can calculate the geometric coefficients required {\color{black} to compute the various terms in the gyrokinetic equation}, as described briefly in Appendix~\ref{app:geometry-appendix1}. For a 3D equilibrium, \GX can read the output data generated by the VMEC code \citep{Hirshman1983}. For a 2D axisymmetric equilibrium, \GX can use a local equilibrium defined using a Miller parametrization~\citep{Miller1998}. The user also has the option to work with simpler geometries like a slab or a cylindrical Z-pinch. Besides these native geometry options, \GX can also read geometry data generated by the geometry module in the GS2 code; for example, this enables the capability to use experimentally-relevant geometry derived from an EFIT equilibrium reconstruction.

%

\section{Fourier-Laguerre-Hermite pseudo-spectral formulation} \label{sec:lh}

Here we present the Fourier-Laguerre-Hermite formulation of the electromagnetic $\delta\!f$ gyrokinetic system. Following \cite{Mandell2018}, and extending to include electromagnetic fluctuations, we solve the gyrokinetic equation by taking a Fourier transform in the perpendicular directions (which we will call $x$ and $y$; the parallel direction will be $z$), a Laguerre transform in $\mu B$, and a Hermite transform in $v_\parallel$. Note that there are no major disadvantages to the choice of $\mu B$ {\it vs.} $\mu$ as a coordinate, and {\color{black} our choice to use $\mu B$ is natural because the Laguerre weight function becomes $\exp(-\mu B)$ just like in the Maxwellian distribution}.

We define the Fourier-Laguerre-Hermite transform of the distribution function $h{\color{black}_s}$ {\color{black} for species $s$} as 
\begin{align}
   \mathcal{L}_\ell \mathcal{H}_m \mathcal{F}_{\bf k_\perp}h_s &=  2\pi  \int_0^\infty d\mu B\, \psi_\ell(\mu B)\int_{-\infty}^\infty d v_\parallel\, \phi_m(v_\parallel) \int dx\, dy\, e^{-i {\bf k_\perp \cdot R}} h_s(x,y,z,v_\parallel, \mu) \notag \\
   &\equiv  H_{\ell,m}^s({\bf k_\perp},z) 
\end{align}
\begin{equation}
 H_{\ell,m}^s({\bf k_\perp},z) = 2\pi  \int_0^\infty d\mu B\, \psi_\ell(\mu B)\int_{-\infty}^\infty d v_\parallel\, \phi_m(v_\parallel) \int dx\, dy\, e^{-i {\bf k_\perp \cdot R}} h_s(x,y,z,v_\parallel, \mu)
\end{equation}
where {\color{black} ${\bf k_\perp} =k_x\nabla x + k_y \nabla y$} is the perpendicular wavenumber vector, $\psi_\ell (\mu B) = (-1)^\ell \Lv{\ell}$ with
\begin{equation}
{\color{black} \mathrm{L}_{\ell}(x)=\frac{e^x}{\ell!}\frac{\mathrm{d}^\ell}{\mathrm{d}x^\ell} x^\ell e^{-x}}
\end{equation}
the Laguerre polynomials, and 
$\phi_m (v_\parallel) = 
{\Hev{m}}/{\sqrt{m!}}$ with 
\begin{equation}
   {\color{black} \mathrm{He}_m(x) = (-1)^m e^{x^2/2}\frac{\mathrm{d}^m}{\mathrm{d}x^m}e^{-x^2/2}}
\end{equation}
the probabilists' Hermite polynomials. In Fourier space, gyroaveraging operations (whether at constant $\bf R$ or constant $\bf r$) simply become multiplications by Bessel functions, so that the Fourier transform of the gyroaveraged potential is
\begin{equation}
    \mathcal{F}_{\bf k_\perp} \gyavgR{\chi} = J_{0s}\Phi({\bf k_\perp},z) -v_{ts}{v_\parallel}J_{0s} A_\parallel({\bf k_\perp},z) + \frac{\tau_s}{Z_s}2\mu B \frac{J_{1s}}{\alpha_s} {\color{black} \frac{\delta\! B_\parallel({\bf k_\perp},z)}{B}}.
\end{equation}
Weighting the gyroaveraged potential by a Maxwellian and Laguerre-Hermite transforming results in
\begin{equation}
  \mathcal{L}_\ell\mathcal{H}_m \mathcal{F}_{\bf k_\perp} \gyavgR{\chi}F_{Ms} = \mathcal{J}_\ell^s \Phi \delta_{m 0} - v_{ts}\mathcal{J}_\ell^s A_\parallel \delta_{m1} + \frac{\tau_s}{Z_s}\left(\mathcal{J}_\ell^s + \mathcal{J}_{\ell-1}^s\right){\color{black} \frac{\delta\!B_\parallel}{B}} \delta_{m0},
\end{equation}
with 
\begin{equation}
    \mathcal{J}_\ell^s \equiv
\int_0^\infty d\mu B\ \psi_\ell J_{0s}(\sqrt{2\mu B b_s})e^{-\mu B} =  \frac{1}{\ell!} \left(-\frac{b_s}{2}\right)^\ell e^{-b_s/2} \qquad (\ell\geq0)
\end{equation}
{\color{black} and $\mathcal{J}_\ell^s\equiv0$ when $\ell<0$ or when $\ell \geq N_\ell$, with $N_\ell$ the number of evolved Laguerre moments.}
The Fourier-Laguerre-Hermite transform of the auxilliary distribution function $g$ is then
\begin{equation}
    \mathcal{L}_\ell \mathcal{H}_m \mathcal{F}_{\bf k_\perp}g_s  \equiv \Gjk^s({\bf k_\perp},z) =  H_{\ell,m}^s - \frac{Z_s}{\tau_s} \mathcal{J}_\ell^s \Phi \delta_{m 0} + \frac{Z_s v_{ts}}{\tau_s}\mathcal{J}_\ell^s A_\parallel \delta_{m1} - \left(\mathcal{J}_\ell^s + \mathcal{J}_{\ell-1}^s\right){\color{black} \frac{\delta\!B_\parallel}{B}} \delta_{m0}.
\end{equation}

The Fourier-Laguerre-Hermite transform of \cref{gkg} then gives
\begin{align}
\pderiv{G_{\ell,m}^s}{t} + \mathcal{N}_{\ell,m}^s & 
  + v_{ts} \nabla_\parallel \left( \sqrt{m+1} \, H^s_{\ell,m+1} + \sqrt{m} \, H^s_{\ell,m-1} \right) 
  \notag \\ & 
  +v_{ts}  \Big[ - (\ell+1) \, \sqrt{m+1} \, H^s_{\ell,m+1} - \ell \, \sqrt{m+1} \, H^s_{\ell-1,m+1} \notag
  \\
  & \qquad\qquad+ \ell \, \sqrt{m} \, H^s_{\ell,m-1} 
  + (\ell+1) \, \sqrt{m} \, H^s_{\ell+1,m-1} \Big] \nabla_\parallel \ln B
  \notag \\ & 
   + i\frac{\tau_s}{Z_s} \omega_{d}^\kappa \Big[ \sqrt{(m+1)(m+2)} \, H^s_{\ell,m+2} + (2m+1)H^s_{\ell,m} +\sqrt{m (m-1)} \, H^s_{\ell,m-2} \Big] \notag\\ &
   + i\frac{\tau_s}{Z_s} \omega_{d}^{\nabla B} \Big[  (\ell+1) \, H^s_{\ell+1,m} + (2\ell+1)H^s_{\ell,m} +\ell \, H^s_{\ell-1,m} \Big] \notag\\ &
   = \mathcal{D}_{\ell,m}^s + \mathcal{C}_{\ell,m}^{ss}.
\label{glmevolve}
\end{align}
Parallel convection, including bounce motion induced by magnetic
trapping in the equilibrium magnetic field, is described by the terms
proportional to $\nabla_\parallel \equiv \bhat \cdot \nabla = (\bhat\cdot\nabla z) \partial/\partial z$. The $\nabla_\parallel H{\color{black}^s_{\ell,m}}$ terms are evaluated spectrally by Fourier transforming in $z$:
\begin{equation}
    \nabla_\parallel H{\color{black}^s_{\ell,m}} \equiv {\color{black} (\bhat\cdot\nabla z)}\mathcal{F}_{k_{\color{black} z}}^{-1}[ i k_{\color{black} z} \mathcal{F}_{k_{\color{black} z}} H{\color{black}^s_{\ell,m}} ],  \label{fftz}
\end{equation}
with
\begin{equation}
    \mathcal{F}_{k_{\color{black} z}}H{\color{black}^s_{\ell,m}} = \int dz\, e^{-i k_{\color{black} z} z} H{\color{black}^s_{\ell,m}}(z).
\end{equation}
{\color{black} Note that a spectral evaluation of this term requires an equispaced grid in $z$, and we also require the $z$ coordinate to be an equal-arc-length coordinate so that $(\bhat\cdot\nabla z)$ is a constant (see \cref{eqarc}) to avoid a convolution in \cref{fftz}.}
Additional details about this operation {\color{black} related to boundary conditions} are given in \cref{sec:bcs}.
Toroidicity gives rise to the terms proportional to the curvature drift operator $i \omega_d^{\kappa} = (1/B)\bhat\times(\bhat\cdot\nabla\bhat)\cdot {i\bf k_\perp}$ and the $\nabla B$ drift operator $i \omega_d^{\nabla B} = (1/B^2)\bhat\times\nabla B\cdot {i\bf k_\perp}$ is the $\nabla B$.
Drive terms from equilibrium gradients, denoted by $\mathcal{D}_{\ell,m}$, are given by
\begin{align}
&\mathcal{D}_{\ell,m=0}^s  =  i \omega_* \, \left[\frac{1}{L_{ns}} \gm{}{\ell}^s +
            \frac{1}{L_{Ts}}  \left[ \ell \gm{}{\ell-1}^s 
            + 2\ell \gm{}{\ell}^s + (\ell+1) \gm{}{\ell+1}^s
               \right] \right]\Phi 
 \notag \\ 
 &\quad +{ \frac{\tau_s}{Z_s}}i \omega_* \, \left[\frac{1}{L_{ns}} [\gm{}{\ell}^s+\gm{}{\ell-1}^s] +
            \frac{1}{L_{Ts}}  \left[ \ell \gm{}{\ell-2}^s + 3\ell\gm{}{\ell-1}^s
            + (3\ell+1) \gm{}{\ell}^s + (\ell+1) \gm{}{\ell+1}^s
               \right] \right]{\color{black} \frac{\delta \!B_\parallel}{B}} 
 \notag \\ 
     &\mathcal{D}^s_{\ell, m=1} = -v_{ts}i\omega_*\left[\frac{1}{L_{ns}}\mathcal{J}_\ell^s + \frac{1}{L_{Ts}}[\ell \mathcal{J}_{\ell-1}^s+(2\ell+1)\mathcal{J}^s_\ell + (\ell+1)\mathcal{J}_{\ell+1}^s] \right]A_\parallel \notag \\
&\mathcal{D}_{\ell,m=2}^s  =  \frac{1}{\sqrt{2}} i\omega_* \, \frac{1}{L_{Ts}}  \,\gm{}{\ell}^s \Phi + { \frac{\tau_s}{Z_s}}\frac{1}{\sqrt{2}} i\omega_* \, \frac{1}{L_{Ts}}  [\gm{}{\ell}^s+\gm{}{\ell-1}^s] {\color{black} \frac{\delta\! B_\parallel}{B}} \notag \\
    &\mathcal{D}_{\ell, m=3}^s = -v_{ts} \sqrt{\frac{3}{2}}i\omega_* \frac{1}{L_{Ts}}\mathcal{J}_\ell^s A_\parallel \notag \\
&\mathcal{D}^s_{\ell, m>3}  = 0 ,
\end{align}
where $i \omega_*\equiv -\nabla x \cdot
\bhat\times i{\bf k_\perp}=i k_y$, and $L_{ns}$ and $L_{Ts}$ are the normalized density and temperature
gradient scale lengths, respectively.

The like-species collision terms are given by
\begin{align}
\mathcal{C}^{ss}_{\ell,0} = &  - \nu_{ss} \, (b_s+2\ell + 0) \, H^s_{\ell,0} \notag\\&+ \nu_{ss} \left(\sqrt{b_s} \, \left(\mathcal{J}_\ell^s + \mathcal{J}_{\ell-1}^s\right) \bar{u}_{\perp s} + 
 2 \left[  \ell \mathcal{J}_{\ell-1}^s + 2 \ell \mathcal{J}_\ell^s +
  (\ell+1)\mathcal{J}_{\ell+1}^s\right] \bar{T}_s \right) , \notag \\
\mathcal{C}^{ss}_{\ell,1} = & - \nu_{ss} \, (b_s + 2 \ell + 1) H^s_{\ell,1} + \nu_{ss} \, \gm{}{\ell}^s \bar{u}_{\parallel s}, \notag \\
\mathcal{C}^{ss}_{\ell,2} = &- \nu_{ss} \, (b + 2\ell + 2) \, H^s_{\ell,2} + \nu_{ss} \, \sqrt{2} \, \mathcal{J}_\ell^s \bar{T}_s, \notag \\
\mathcal{C}^{ss}_{\ell,m}=&- \nu_{ss} (b_s + 2 \ell +m) H^s_{\ell,m}, \qquad \qquad (m>2) \label{colls}
\end{align}
 with the field-particle terms given by
\begin{gather}
\bar{u}_{\parallel s} = \int d^3{\bf v} J_{0s} v_\parallel h_s = \sum_{\ell=0}^{\color{black} N_\ell} \gm{}{\ell}^s H_{\ell,1}^s, \label{eq:ubarHL}\\
\bar{u}_{\perp s} = \int d^3{\bf v} J_{1s} v_\perp h_s = \sqrt{b_s} \sum_{\ell=0}^{\color{black} N_\ell}
 \left(\mathcal{J}_\ell^s+\mathcal{J}_{\ell-1}^s\right)H^s_{\ell,0}, \label{eq:uperpHL}\\
\bar{T}_{\parallel s} = \int d^3{\bf v} J_{0s} \left(v_\parallel^2-1\right) h_s = \sqrt{2} \sum_{\ell=0}^{\color{black} N_\ell} \gm{}{\ell}^s H^s_{\ell,2}, \label{eq:tparbarHL}\\
\bar{T}_{\perp s} = \int d^3{\bf v} J_{0s} \left(\mu B-1\right) h_s = \sum_{\ell=0}^{\color{black} N_\ell} \left[\ell \mathcal{J}^s_{\ell-1}+2\ell\mathcal{J}^s_{\ell} +(\ell+1)\mathcal{J}^s_{\ell+1} \right]H_{\ell,0}^s,
\label{eq:tbarHL} \\
{\color{black} \bar{T}_s = \frac{1}{3}(\bar{T}_\parallel + 2 \bar{T}_\perp)}
\end{gather}

The nonlinear terms are evaluated pseudo-spectrally in $(x,y,z,\mu B, m)$ space to avoid convolutions in both Fourier and Laguerre coefficients as \citep{Mandell2018},
\begin{align}
    \mathcal{N}_{\ell,m}^s &= \mathcal{L}_\ell \mathcal{H}_m \mathcal{F}_{\bf k_\perp} [\gyavgR{\vEM}\cdot\nabla {\color{black} h_s}],
\end{align}
where the complete pseudo-spectral expression for this term is given in \cref{app:nl}.

Taking the Fourier-Laguerre-Hermite transform of the field equations, the quasineutrality equation, \cref{qneuts}, becomes
\begin{align}
    \sum_s \frac{Z_s^2 n_s}{\tau_s}\Phi =  \sum_s Z_s n_s \int\,d^3{\bf v}\, J_{0s} h_s = \sum_s Z_s n_s \sum_{\ell=0}^{\color{black} N_\ell} \mathcal{J}_\ell^s H_{\ell,0}^s,
\end{align}
the parallel component of Amp\`ere's law, \cref{ampere}, becomes
\begin{align}
    k_\perp^2 A_\parallel =  \frac{\beta_\mathrm{ref}}{2}\sum_s Z_s n_s v_{ts} \int d^3{\bf v}\, v_\parallel J_{0s} h_s= \frac{\beta_\mathrm{ref}}{2}\sum_s Z_s n_s v_{ts} \sum_{\ell=0}^{\color{black} N_\ell} \mathcal{J}_\ell^s H^s_{\ell, 1},
\end{align}
and the perpendicular component of Amp\`ere's law, \cref{dB}, becomes
\begin{align}
    {\color{black} \frac{\delta\! B_\parallel}{B}}=  -\frac{\beta_\mathrm{ref}}{2{\color{black} B^2}}\sum_s n_s \tau_s \int d^3\vec{v}\ 2\mu B \frac{J_{1s}}{\alpha_s} h_s = -\frac{\beta_\mathrm{ref}}{2{\color{black} B^2}}\sum_s n_s\tau_s \sum_{\ell=0}^{\color{black} N_\ell} (\mathcal{J}^s_\ell+\mathcal{J}^s_{\ell-1}) H^s_{\ell, 0}.
\end{align}
To solve these field equations numerically, it is more convenient to express them in terms of $G$ rather than $H$, which results in
\begin{gather}
    \sum_s \frac{ Z_s^2n_s}{\tau_s}\left(1-\sum_{\ell=0}^{\color{black} N_\ell}(\mathcal{J}^s_\ell)^2\right)\Phi - \sum_s Z_s n_s \sum_{\ell=0}^{\color{black} N_\ell}\mathcal{J}^s_\ell(\mathcal{J}^s_\ell+\mathcal{J}^s_{\ell-1}){\color{black} \frac{\delta\! B_\parallel}{B}} = \sum_s Z_s n_s \sum_{\ell=0}^{\color{black} N_\ell} \mathcal{J}^s_\ell G^s_{\ell,0} \\
    \left(k_\perp^2 + \frac{\beta_\mathrm{ref}}{2}\sum_s \frac{Z_s^2 n_s}{m_s} \sum_{\ell=0}^{\color{black} N_\ell} (\mathcal{J}^s_\ell)^2\right)A_\parallel = \frac{\beta_\mathrm{ref}}{2}\sum_s Z_s n_s v_{ts} \sum_{\ell=0}^{\color{black} N_\ell} \mathcal{J}^s_\ell G^s_{\ell,1} \\
     \frac{\beta_\mathrm{ref}}{2{\color{black} B^2}}\sum_s Z_s n_s \sum_{\ell=0}^{\color{black} N_\ell}\mathcal{J}^s_\ell(\mathcal{J}^s_\ell+\mathcal{J}^s_{\ell-1})\Phi + \left(1 + \frac{\beta_\mathrm{ref}}{2{\color{black} B^2}}\sum_s n_s \tau_s \sum_{\ell=0}^{\color{black} N_\ell}(\mathcal{J}^s_\ell + \mathcal{J}^s_{\ell-1})^2\right){\color{black} \frac{\delta\! B_\parallel}{B}} \qquad\qquad\notag \\
     \qquad\qquad\qquad\qquad\qquad\qquad \qquad\qquad\qquad= -\frac{\beta_\mathrm{ref}}{2{\color{black} B^2}}\sum_s n_s \tau_s \sum_{\ell=0}^{\color{black} N_\ell}(\mathcal{J}^s_\ell + \mathcal{J}^s_{\ell-1})G^s_{\ell,0}.
\end{gather}
{\color{black} Note that while the sum $\sum (\mathcal{J}_\ell^s)^2$ could be computed analytically in the $N_\ell\rightarrow \infty$ limit as $\Gamma_0(b) = I_0(b) e^{-b}$, with $I_0$ the modified Bessel function of the first kind, doing so breaks the energetic consistency of the equations \citep{Mandell2018}; thus these terms should be evaluated as truncated sums.}

\subsection{Hyper-dissipation and closure} \label{sec:hyper}

Grid-scale dissipation terms are used to ensure numerical stability and robustness. For dissipation in configuration space, we employ a hyper-viscosity term of the standard form
\begin{equation}
\left(\pderiv{G_{\ell,m}}{t}\right)_\textrm{hyper-viscosity} = - D \left(\frac{k_\perp^2}{k_{\perp,\mathrm{max}}^2}\right)^n G_{\ell,m},
\end{equation}
where {\color{black} $k_{\perp,\mathrm{max}}$ is the largest dealiased perpendicular wavenumber in the simulation, and} typical values of the variable parameters for nonlinear simulations are $D=0.05$ and $n=2$. This provides a sink at the shortest wavelengths in the domain, which can be useful in nonlinear simulations to avoid spectral pileup. {\color{black} Unlike in other Eulerian gyrokinetic codes which include dissipation along the parallel direction via upwinding, \citep{kotschenreuther1995,jenko2001,Candy2016,Barnes2019}, the spectral discretization scheme used for the parallel direction in \GX does not itself introduce dissipation in the parallel direction. The hypercollision operator and the parallel boundary conditions discussed below introduce some parallel dissipation. A standard parallel hyper-dissipation operator of the form $k_\parallel^{2n}$ could also be included, but we find that this is not necessary for the linear and nonlinear cases that we consider.} 

Fine-scale structure in velocity space must also be regulated. 
The Dougherty collision operator fulfills this purpose by acting increasingly strongly
on higher Laguerre and Hermite moments, limiting their amplitude. Thus for a given
collisionality, there is a physical cutoff at some ${\color{black} \ell_\mathrm{cut}}$ and ${\color{black}m_\mathrm{cut}}$ beyond which fine scales
in velocity space are completely wiped out by collisions. At this point, closure of the Laguerre-Hermite moment series by simple truncation (setting $G_{\ell,m}=0$ for $\ell\geq{\color{black} \ell_\mathrm{cut}}$ or $m\geq{\color{black} m_\mathrm{cut}}$) is justified. This is the simplest high-resolution closure, but not the only
option. One can also obtain an asymptotically correct collisional closure by assuming
the collision term becomes dominant in the unresolved moment equations {\color{black} \citep{Zocco2015,Loureiro2016,Jorge2017,Frei2022}}.

In the limit of low collisionality or lower Laguerre–Hermite resolution, however, the closure situation is more
complicated. Unresolved moments are not expected to be negligible at collisionalities
of interest, so closure by truncation will generally give poor results.
One possible approach is to follow the collisionless closure approach pioneered by \citet{Hammett1990}, which was used and extended in the gyrofluid models
of \citet{Dorland1993, Beer1996, Snyder2001, Smith1997} to model parallel, toroidal, and nonlinear FLR phase mixing. 
However, the development of generalized closures in toroidal geometry for a system with an arbitrary number of moments is complicated and beyond the scope of the current work. 
{\color{black}Instead, we have found that employing a hyper-collision operator that provides a sink at large $m$ allows well-behaved results even with simple closure by truncation at relatively low resolution. The operator takes the generic form
\begin{equation}
\left(\pderiv{G_{\ell,m}}{t}\right)_\textrm{hyp-coll}  =
-\nu_\mathrm{hyp} m^{p} G_{\ell, m} \quad (m>2).
\end{equation}
We ensure that the operator conserves density, momentum, and energy by only operating on moments with $m>2$. Similar hyper-collision operators (in Hermite space only) have been used in other contexts \citep{Parker2015, Parker2015a, Loureiro2016}, and \citet{Parker2015a} also develops theory of a `hypercollisional plateau' where the behavior of the operator is insensitive to the choice of the variable parameters.
In our operator, the coefficient $\nu_\mathrm{hyp}$ is chosen to approximately remove the same amount of fluctuation energy from the largest $m$'s as would be removed by parallel phase mixing. The details are provided in \cref{app:hyper}, and the result is
\begin{equation}
     \nu_\mathrm{hyp} = 2.5 f_\mathrm{hyp} \frac{p+1/2}{m_\mathrm{max}^{p+1/2}}|k_\parallel|v_{ts},
\end{equation}
where $f_\mathrm{hyp}$ is an adjustable coefficient that is typically set to unity.
The $|k_\parallel|$ operator is evaluated spectrally by Fourier transforming in $z$.
The exponent $p$ is chosen to be $p=N_m/2$, which ensures that only the highest several Hermite modes are strongly damped, roughly independent of $N_m$. While one could consider a similar hyper-dissipation mechanism for large $\ell$, we find that Laguerre hyper-collisions are unnecessary even when using relatively low Laguerre resolution ($N_\ell\sim 4$) for problems of interest because  phase mixing in $v_\parallel$ dominates. 
}

\subsection{\color{black} Boundary conditions} \label{sec:bcs}

In the flux tube approach we assume statistical periodicity in the perpendicular $(x,y)$ plane. In the parallel direction, we use the ``twist-and-shift'' boundary condition \citep{Beer1995}, which in the spectral representation can be accomplished by linking modes with different $k_x$ values into an extended $z$ domain, such that
\begin{equation}
    {\color{black} H_{\ell,m}(k_x, k_y, z) = H_{\ell,m}(k_x+\delta k, k_y, z+2\pi)}
\end{equation}
with $\delta k = 2\pi k_y \hat{s}$. 
We can then perform a Fourier transform on each extended $z$ domain to evaluate the parallel streaming terms as given by \cref{fftz}.

A Fourier representation naturally enforces a periodic boundary condition on $h$ at the ends of each extended domain, but this is not the physically correct boundary condition for non-zonal modes, which should have a zero incoming boundary condition {\color{black} in the infinite ballooning limit. Following \citet{kotschenreuther1995}, we would like to approximate this by enforcing a zero incoming boundary condition on $h$ at the ends of the extended domain, but this boundary condition cannot be enforced directly in Hermite space. Instead, we use a wave-absorbing layer} \citep[see \emph{e.g.}][\color{black} and references within]{durran2010a} near the ends of each extended $z$ domain for the non-zonal components of $h$ to damp out perturbations {\color{black} leaving and} re-entering the domain {\color{black} due to the natural periodicity of the Fourier representation}. Following \citet{beer1995a}, we use a damping filter that is zero in most of the domain and ramps up smoothly near the ends of the extended domain as $d(\hat{z}) = A[1-2\hat{z}^2/(1+\hat{z}^4)]$, where $\hat{z} = (z-z_\mathrm{end})/z_\mathrm{width}$ is a normalized distance from the ends $z_\mathrm{end}$. The width $z_\mathrm{width}$ and the scaling factor $A$ are adjustable parameters; {\color{black} in \cref{app:damp} we have performed a convergence study and found} $z_\mathrm{width} = L_z/8$ and {\color{black} $A = 0.1/\Delta t$} to work well, where $L_z$ the total length of the extended domain. The wave-absorbing damping operation can then be expressed as
\begin{equation}
    \left(\pderiv{G_{\ell, m}}{t}\right)_\mathrm{ends} = -d(\hat{z}) H_{\ell,m}. \qquad \qquad (|\hat{z}|<1,\  k_y \neq 0)
\end{equation}
The zonal modes $(k_y=0)$ are not filtered because these modes should be physically periodic, so the natural periodic boundary condition from the Fourier representation is the correct one. {\color{black} We have found this wave-absorbing layer to be particularly necessary for simulations with kinetic electrons.}

The standard twist-and-shift boundary condition can be ill-suited for low magnetic shear regions, particularly in stellarators. \citet{Martin2018} have developed a generalization of the twist-and-shift boundary condition that alleviates some of these issues by using the integrated local shear (rather than global shear) in the `twist'. {\color{black} This generalized twist-and-shift boundary condition can be used in GX for stellarator geometries with stellarator symmetry}. Investigation of a non-twisting flux-tube boundary condition \citep{Ball2021} using GX is also in progress. 

{\color{black}
\citet{Martin2018} prescribe methods to choose the parallel domain length to enforce exact periodicity or continuous magnetic drifts at the ends of the flux tube. In addition to these options, we adopt an additional criteria for choosing the parallel domain length: to enforce a desired aspect ratio for the perpendicular computational domain. When using the standard twist-shift boundary condition, the aspect ratio is quantized as $L_x/L_y = J/(2\pi|\hat{s}|)$, with $J$ a non-zero integer. This results in $L_x \propto L_y/\hat{s}$, which makes resolution requirements challenging for $\hat{s}\ll 1$. In the generalized twist-shift boundary condition of \citet{Martin2018}, the perpendicular aspect ratio of the flux tube is given by
\begin{equation}
    \frac{L_x}{L_y} = J\left[\frac{|\nabla x|^2}{2|\nabla x\cdot \nabla y|}\right]_{z_\mathrm{end}},
\end{equation}
where $z_\mathrm{end}$ is the end of the flux tube and $J$ is again a non-zero integer. In a stellarator, this quantity varies as a function of the parallel domain length (see Fig. 7 of \citep{Martin2018}). 
The generalized twist-shift boundary condition thus provides the freedom to choose $z_\mathrm{end}$ to obtain the desired aspect ratio (for some integer $J$). Using this prescription to enforce an aspect ratio of order unity can then alleviate the resolution requirements for geometries with small magnetic shear.
}

%
%
%

\section{Timestepping schemes} \label{sec:time}
\GX uses explicit time integration methods to advance the system.
Along with several standard Runge-Kutta (RK) and strong-stability-preserving Runge-Kutta (SSP RK) schemes \citep{Gottlieb2001}, we also provide an option for the SSP ten-stage fourth-order method of \citet{Ketcheson2008} (a low-storage method ideal for the memory constraints of GPU computing). 
Additionally, the sparse banded structure of the linear system of \cref{glmevolve} and the fact that the field equations only involve the lowest order moments (instead of requiring integration over all velocity space) could allow the development of efficient implicit-explicit (IMEX) schemes for electron dynamics that avoid the timestep restriction of the fast electron motion; this will be a topic for future work. {\color{black} We have used the standard RK3 (third-order) scheme for all of the benchmark calculations presented in \cref{sec:benchmarks}}. 

In all of our timestepping schemes the timestep size $\Delta t$ is constrained by {\color{black} the stability region of the scheme. While in principle the linear eigenvalues of the system could be computed exactly and used to constrain the timestep to ensure that all eigenvalues lie within the stability region, we instead estimate the maximum (linear and nonlinear) frequencies on the grid in each of the $(x,y,z)$ directions via
\begin{gather}
    \omega_{x,\mathrm{max}} \approx \max\left(k_{x,\mathrm{max}} (\frac{\tau_s}{Z_s}\vec{v}_d\cdot\nabla x)_\mathrm{max},\ k_{x,\mathrm{max}} (\vec{v}_E\cdot \nabla x)_\mathrm{max}\right) \\
    \omega_{y,\mathrm{max}} \approx \max\left(k_{y,\mathrm{max}} (\frac{\tau_s}{Z_s}\vec{v}_d\cdot\nabla y)_\mathrm{max},\ k_{y,\mathrm{max}} (\vec{v}_E\cdot \nabla y)_\mathrm{max}\right) \\
    \omega_{z,\mathrm{max}} \approx \max \left((\bhat\cdot\nabla z)k_{z,\mathrm{max}} v_{\parallel,\mathrm{max}}  v_{t,\mathrm{max}},\  \omega_{A,\mathrm{max}}\right)
\end{gather}
with 
\begin{equation}
   \omega_{A,\mathrm{max}} = \frac{(\bhat\cdot\nabla z)k_{z,\mathrm{max}} v_{te}}{\sqrt{\frac{\beta_\mathrm{ref}n_e \tau_e}{2}\frac{m_i}{m_e} + k_{\perp,\mathrm{min}}^2 \rho_s^2}}
\end{equation}
the maximum shear Alfv\'en frequency on the grid. In the electrostatic limit ($\beta_\mathrm{ref} = 0$) this reduces to the high-frequency $\omega_H$ mode \citep{lee1987}.
Here, $k_{z,\mathrm{max}}$ is the largest parallel wavenumber in the simulation and $k_{\perp,\mathrm{min}}$ is the smallest perpendicular wavenumber. To evaluate $v_{\parallel,\mathrm{max}}$ and $(\mu B)_\mathrm{max}$ we use the maximum value of the Gauss-Hermite and Gauss-Laguerre collocation grids, respectively, which depend on the number of spectral modes used in the calculation, $N_m$ and $N_\ell$. 

The constraint on the maximum stable timestep is then approximately given by
\begin{equation}
    (\omega_{x,\mathrm{max}} + \omega_{y,\mathrm{max}} + \omega_{z,\mathrm{max}})\Delta t \lesssim C s,
\end{equation}
where $s$ is dependent on the stability region for the particular timestepping scheme (for example, $s=1.73$ for RK3, and $s=2.82$ for RK4), and $0<C\leq 1$ is a user-defined scaling factor. Formally, the stability constraint with $C=1$ is only valid for linear modes that do not grow or decay; thus we typically use $C\sim0.5-0.9$ to ensure stability in the full nonlinear system with 
 a spectrum of growing and damped modes.}

\section{Numerical benchmarks} \label{sec:benchmarks}
In this section we show a variety of linear and nonlinear benchmarks of \GX by comparing results with widely-benchmarked gyrokinetic codes (GS2,  stella, GENE) in both tokamak and stellarator geometry. {\color{black} In all cases, both linear and nonlinear, an initial value problem is solved.} The numerical resolution and associated parameters used in each case are given in \cref{app:resolution}.

\begin{figure}
    \centering
    \includegraphics[width=.6\textwidth]{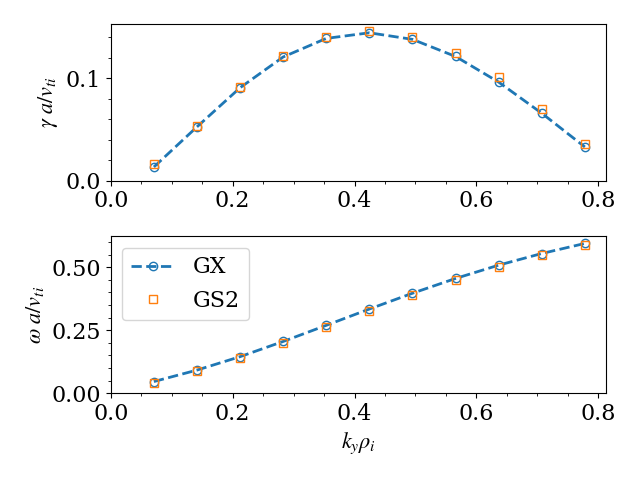}
    \caption{Normalized growth rates (top) and real frequencies (bottom) as a function of the normalized binormal wavenumber $k_y{\color{black} \rho_i}$ from \GX (blue circles and dotted lines) and GS2 (yellow squares) for {\color{black} CBC} parameters using a Boltzmann electron response.}
    \label{fig:lin_adiab}
\end{figure}

\subsection{Tokamak benchmarks: Cyclone base case 
}

For benchmarks in tokamak geometry, we choose a configuration based on the `Cyclone base case' (CBC), a widely-used benchmark case with concentric circular flux surfaces \citep{Dimits2000}. For this we use a Miller local equilibrium \citep{Miller1998} with $R_0/a=2.78$, $r/a=0.5$, $q=1.4$, $\hat{s}=0.8$, $\kappa=1.0$, $\delta=0.0$. Unless otherwise noted, the normalized gradient scale lengths are taken to be $a/L_{ni} = a/L_{ne} = 0.8$, $a/L_{Ti} = a/L_{Te}= 2.49$, and we also take $T_i = T_e$. {\color{black} In cases with kinetic electrons we take $m_i/m_e = 3670$ (deuterium ions).} This series of benchmarks is inspired by the set of tokamak benchmarks chosen for stella by \citet{Barnes2019}. 

\subsubsection{Linear results}


We first make a linear comparison taking a Boltzmann (adiabatic) electron response, so that the quasineutrality equation \labelcref{qneuts} reduces to
\begin{equation}
 \int d^3{\bf v}\ \gyavgr{h_s} =  \sum_{\ell=0}^{\color{black} N_\ell}\mathcal{J}_\ell H_{\ell,0}  =\frac{T_e}{T_i}[\Phi - \langle\langle \Phi\rangle\rangle] + \Phi,
 \label{qneuta}
\end{equation}
where $\langle\langle\Phi\rangle\rangle$ denotes the flux-surface average of $\Phi$.
In \cref{fig:lin_adiab} we show normalized growth rates (top panel) and real frequencies (bottom) as a function of the normalized binormal wavenumber $k_y \rho_i$. \GX results are shown with open blue circles connected by dashed lines, while results from GS2 are shown with yellow squares. The agreement between the two codes is very good across the range of unstable $k_y$ values. Here we included a small amount of collisions in both codes to provide velocity-space regularization, taking the normalized ion collision frequency to be $\nu_{ii} = 10^{-2} \, v_{ti}/a$. Hyper-collisions are not used in \GX in this case.

\begin{figure}
    \centering
    \includegraphics[width=.6\textwidth]{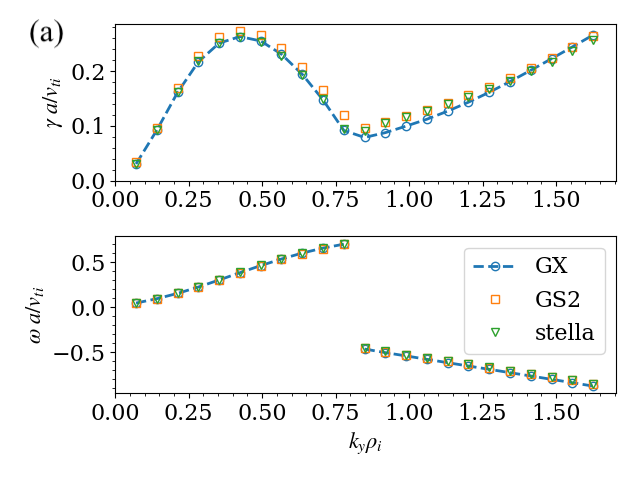}
    \includegraphics[width=.6\textwidth]{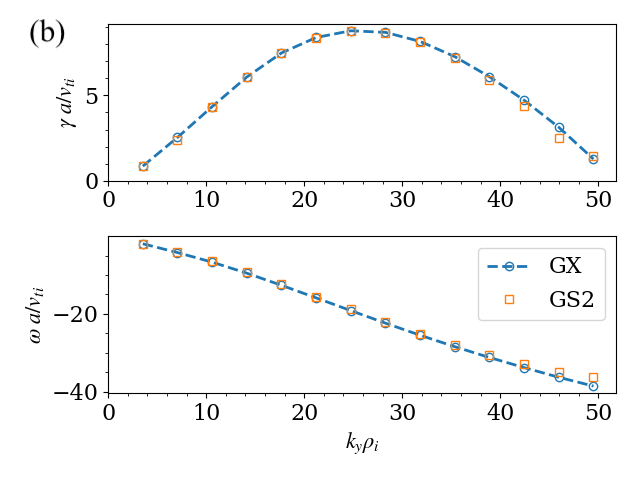}
    \caption{Normalized growth rates (top) and real frequencies (bottom) as a function of the normalized binormal wavenumber $k_y{\color{black} \rho_i}$ from \GX (blue circles and dotted lines), GS2 (yellow squares), and stella (green inverted triangles) for {\color{black} CBC} parameters using a kinetic electron response in the electrostatic limit ($\beta_\mathrm{ref} = 10^{-5}$). Ion scales (ITG and TEM) are shown in (a) and electron scales (ETG) are shown in (b).}
    \label{fig:lin_kin}
\end{figure}

\begin{figure}
    \centering
    \includegraphics[width=.6\textwidth]{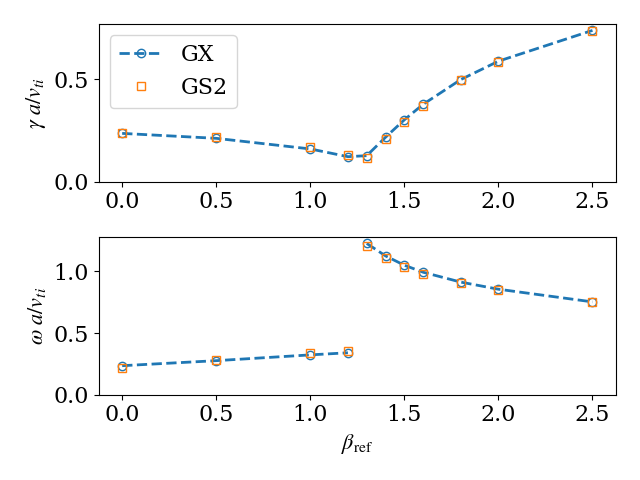}
    \caption{Normalized growth rates (top) and real frequencies (bottom) as a function of the reference beta $\beta_\mathrm{ref}$ from \GX (blue circles and dotted lines) and GS2 (yellow squares) for {\color{black} CBC}  parameters using a kinetic electron response, taking $k_y \rho_i = 0.3$. Both codes show the transition from ion-temperature-gradient (ITG) instability at low $\beta_\mathrm{ref}$ to kinetic-ballooning-mode (KBM) instability at high $\beta_\mathrm{ref}$.}
    \label{fig:itg_kbm}
\end{figure}

We next make linear comparisons using kinetic electrons between \GX, GS2, and stella. \cref{fig:lin_kin}a shows normalized growth rates and real frequencies as a function of $k_y \rho_i$ at ion scales in the electrostatic limit, with $\beta_\mathrm{ref}=10^{-5}$ (we retain electromagnetic fluctuations even at low $\beta$ to alleviate the timestep restriction from the electrostatic $\omega_H$ mode \citep{lee1987}). 
In this case we take $\nu_{ee} = 10^{-2}  \, v_{ti}/a$ and $\nu_{ii} = 1.65\times 10^{-4} \, v_{ti}/a$; for best comparison with \GX, we use the stella's Dougherty collision model option, even though stella also includes a more accurate Fokker-Planck collision operator (which produces results that agree even more closely with GS2 in this case). Hyper-collisions were again not used in \GX. We observe excellent agreement between \GX and stella for the ITG branch $(\omega>0)$, but the agreement is not as good for the trapped-electron-mode (TEM) branch $(\omega<0)$. If there is a weakness in our scheme it is accurately capturing linear TEM instability, which requires resolving sharp trapped-passing boundaries in velocity space (which is not optimal in $v_\parallel,\mu$ coordinates) and is sensitive to the details of the collision operator (our Dougherty model collision operator is not as accurate as the collision operators in other standard gyrokinetic codes like GS2). Indeed, even getting this level of agreement required 128 Hermite modes in \GX (for more resolution details, see \cref{app:resolution}). A more detailed study of TEM modes is left to future work, where we will explore more accurate collision operators and methods to enhance convergence. {\color{black} Note that \citet{Frei2021,Frei2022a} have demonstrated the success of the Laguerre-Hermite method in resolving TEM modes with a more accurate collision operator.} We also show growth rates and frequencies as a function of $k_y \rho_i$ at electron scales in \cref{fig:lin_kin}b, again in the electrostatic limit. There is once again strong agreement between \GX and GS2 for these electron-temperature-gradient (ETG) modes.

We finally perform an electromagnetic linear benchmark of the transition from ITG instability to kinetic ballooning mode (KBM) instability.  In this benchmark we include perpendicular magnetic fluctuations via finite $A_\parallel$ but we neglect parallel magnetic fluctuations ($\delta\!B_\parallel=0$). Taking $k_y\rho_i=0.3$, \cref{fig:itg_kbm} shows normalized growth rates and real frequencies as a function of $\beta_\mathrm{ref}$, the reference beta. As $\beta_\mathrm{ref}$ increases, both \GX and GS2 agree well and show  moderate stabilization of the ITG mode until the transition to KBM around $\beta_\mathrm{ref} = 1.3\%$. {\color{black} Additional work studying reactor-relevant electromagnetic instabilities (\emph{e.g.} micro-tearing and toroidal Alfv\'en eigenmodes) with \GX will be reported in future publications; successful modeling of collisionless micro-tearing with a Laguerre-Hermite formulation has been shown by \citet{Frei2023}.}

\begin{figure}
    \centering
    \includegraphics[width=.6\textwidth]{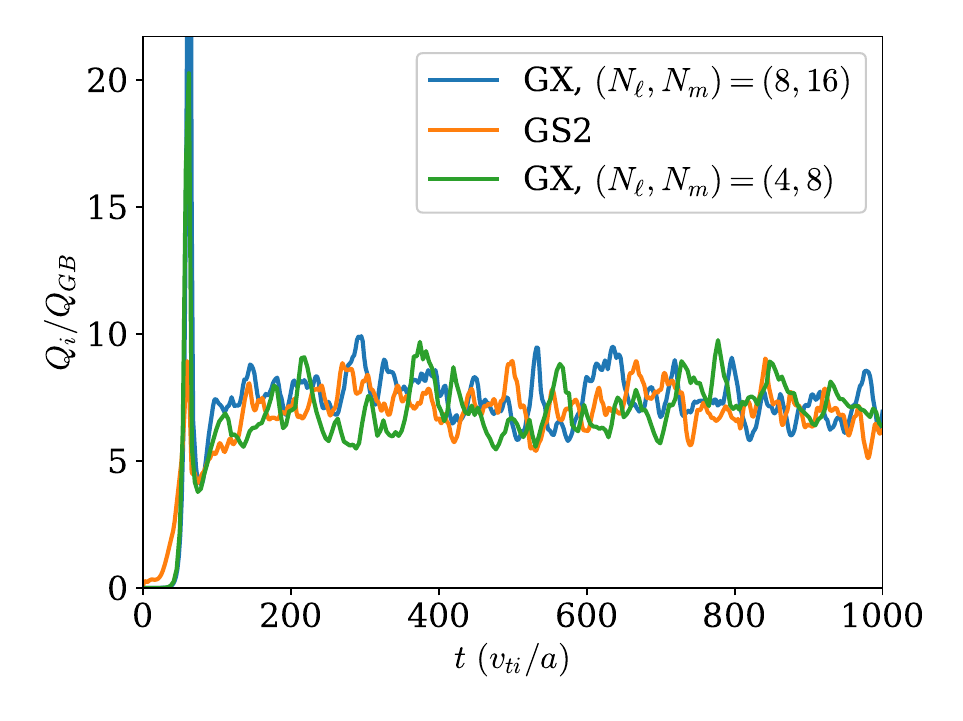}
    \includegraphics[width=.6\textwidth]{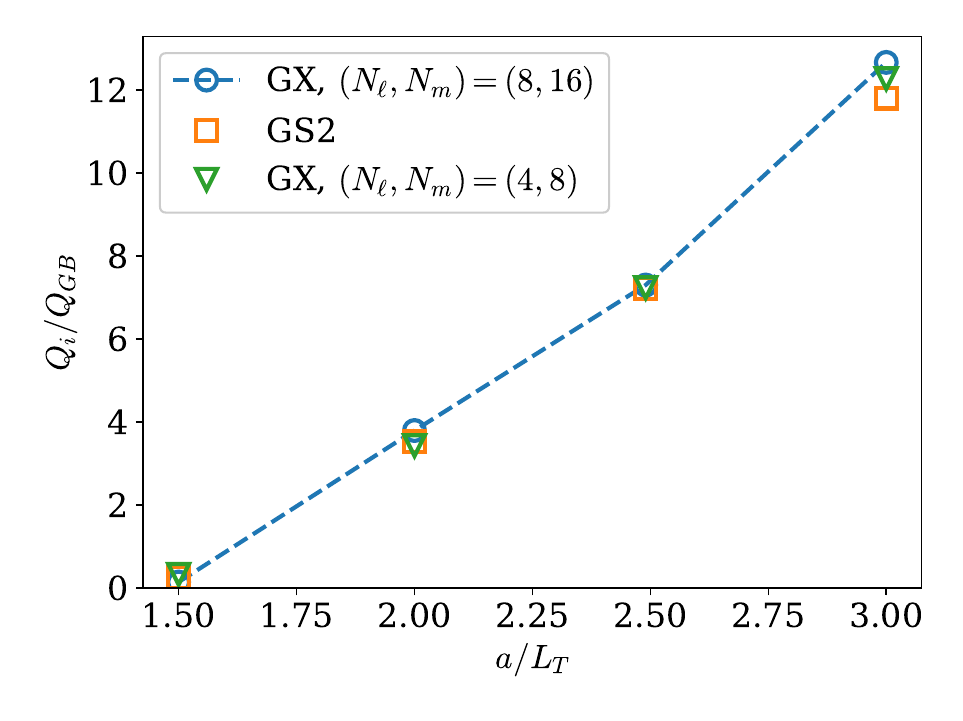}
    \caption{Left: Time traces of ion heat flux $Q_i$, normalized to gyro-Bohm units, $Q_{GB}= n_i T_i v_{ti}\rho_i^2/a^2$, for the CBC with a Boltzmann electron response from \GX with  moderate (blue) and coarse (green) velocity resolution, and {\color{black} GS2 (yellow)}. Right: Time-averaged gyro-Bohm-normalized ion heat flux, $Q_i/Q_{GB}$, as a function of the normalized inverse ion temperature gradient scale length, $a/L_{Ti}$.}
    \label{fig:nl_cyclone}
\end{figure}

\subsubsection{Nonlinear results} \label{sec:nl_cyclone}

We first perform nonlinear CBC calculations using a Boltzmann electron response. \cref{fig:nl_cyclone}a shows time traces of the ion heat flux in gyro-Bohm units {\color{black} (computed using the expressions in \cref{app:fluxes})}, showing very good agreement in the time-averaged heat flux amongst \GX and GS2.  For \GX, we show a case with moderate velocity resolution, $( N_\ell, N_m)=(8,16)$, along with a coarse resolution case with $( N_\ell, N_m)=(4,8)$. Additional details about velocity-space convergence for this case are given in \cref{sec:convergence}. We also show in \cref{fig:nl_cyclone}b a scan of the ion temperature gradient parameter, $a/L_{Ti}$, which again shows excellent agreement between \GX and GS2 for all cases at both resolutions. The agreement at $a/L_T=1.5$ is particularly notable because this is in the Dimits shift regime, where the system is linearly unstable to ITG but turbulence is suppressed by zonal flow dynamics. This regime was especially troublesome for the Beer gyrofluid model \citep{Beer1996,Dimits2000}, but we see that by extending that model to more moments (and neglecting the collisionless closure schemes) via our Laguerre-Hermite approach we can recover the Dimits shift {\color{black} with $(N_\ell,N_m)=(4,8)$. \citet{Hoffmann2023a} have studied the convergence of the Laguerre-Hermite basis for this case and also found that $(N_\ell,N_m)\geq(4,8)$ is required to accurately capture the Dimits shift. Further from marginality, \citet{Hoffmann2023a} have found that even lower velocity resolution can be used with reasonable accuracy, which is consistent with the convergence study in \cref{sec:convergence}.} Here, both GX and GS2 included a small amount of collisions, taking the normalized ion collision frequency to be $\nu_{ii} = 10^{-2} \, v_{ti}/a$. Hyper-collisions were included in the \GX calculations, with details given in \cref{app:resolution_nl_cyclone}.

\begin{figure}
    \centering
    \includegraphics[width=.6\textwidth]{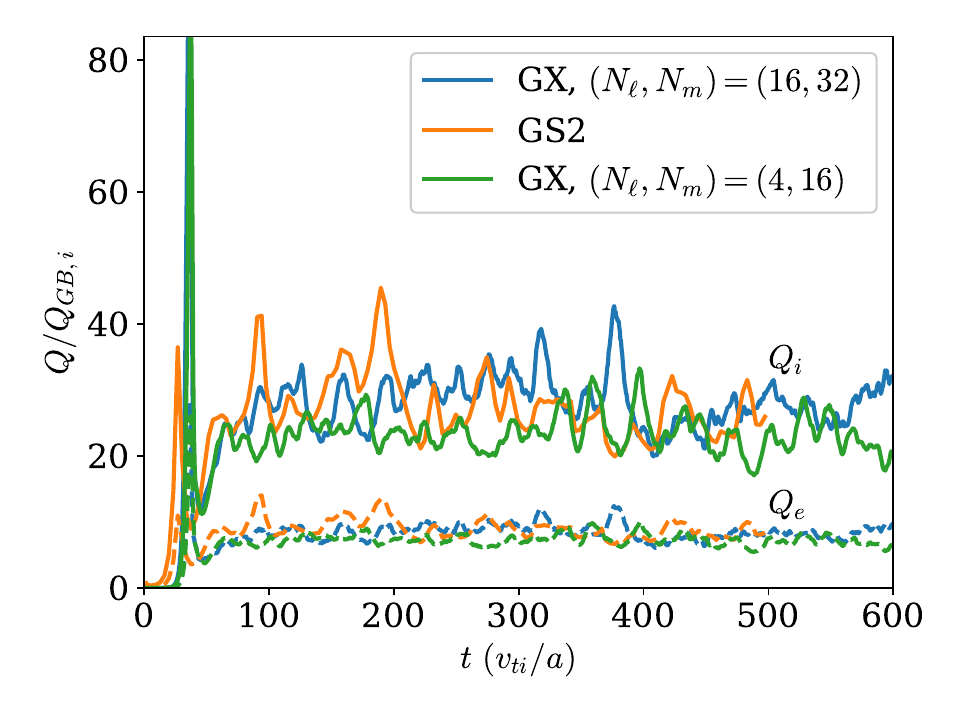}
    \caption{Time traces of ion (solid) and electron (dashed) heat flux, normalized to ion gyro-Bohm units, for the CBC with kinetic electrons from {\color{black} \GX with high velocity resolution (blue) and moderate velocity resolution (green), and GS2 (yellow)}.}
    \label{fig:nl_cyclone_kin}
\end{figure}

Nonlinear calculations with kinetic electrons are presented in \cref{fig:nl_cyclone_kin}, which shows excellent agreement amongst \GX and GS2 for the time average of the heat flux in both the ion and electron channels. For \GX, we show a case with moderate velocity resolution, $( N_\ell, N_m)=(4,16)$, along with a higher resolution case with $( N_\ell, N_m)=(16,32)$. Additional details about velocity-space convergence for this case are given in \cref{sec:convergence}. Here we took {\color{black} $\beta_\mathrm{ref} = 10^{-3}$ along with} $\nu_{ee} = 10^{-2}  \, v_{ti}/a$ and $\nu_{ii} = 1.65\times 10^{-4} \, v_{ti}/a$ for collisions in all codes, and hyper-collisions were again used in \GX (see \cref{app:resolution_nl_cyclone_kin}).

\subsection{Stellarator benchmarks: W7-X}

For benchmarks in stellarator geometry, we choose a magnetic configuration from W7-X that has recently been used by \citet{Gonzalez-Jerez2022} to benchmark stella and GENE. The numerical equilibrium is obtained from VMEC \citep{Hirshman1983}, and we choose the so-called bean flux tube with $\alpha_0=0$. To study ITG-driven instabilities and turbulence we take $a/L_{Ti}=3$ and $a/L_n=1$. {\color{black} The generalized twist-and-shift boundary condition of \citet{Martin2018} is used.}

\begin{figure}
    \centering
    \includegraphics[width=.6\textwidth]{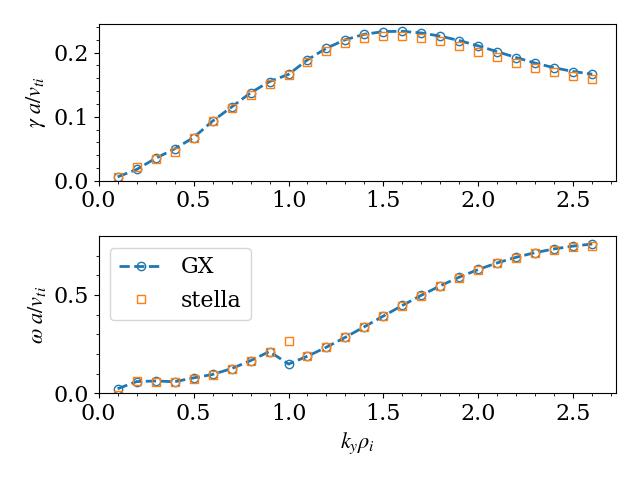}
    \caption{Normalized growth rates (top) and real frequencies (bottom) as a function of the normalized binormal wavenumber $k_y$ from \GX (blue circles and dotted lines) and stella (yellow squares) for W7-X  bean flux-tube geometry and ITG parameters using a Boltzmann electron response.}
    \label{fig:lin_w7x}
\end{figure}

\begin{figure}
    \centering
    \includegraphics[width=.6\textwidth]{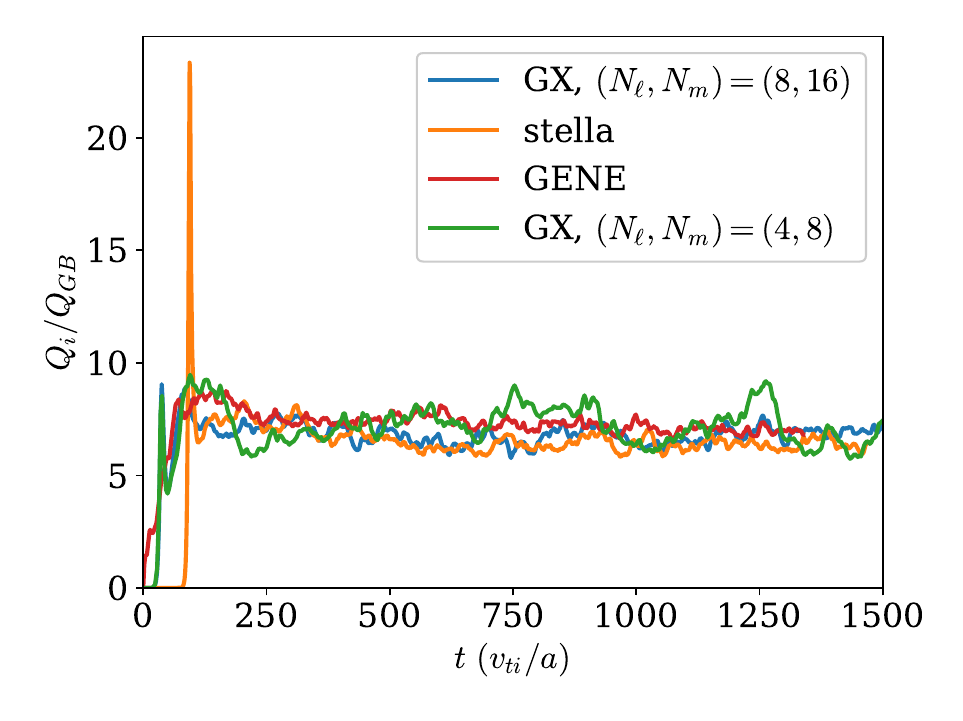}
    \caption{Time traces of ion heat flux $Q_i$, normalized to gyro-Bohm units, $Q_{GB}= n_i T_i v_{ti}\rho_i^2/a^2$, for the W7-X bean flux-tube geometry with a Boltzmann electron response. Results are shown from \GX with  moderate (blue) and coarse (red) velocity resolution, stella (yellow), and GENE (green).}
    \label{fig:nl_w7x}
\end{figure}

\subsubsection{Linear results}
We first make a linear comparison between \GX and stella, taking a Boltzmann electron response. \cref{fig:lin_w7x} shows the normalized growth rates (top) and real frequencies (bottom) for a range of $k_y$ modes, with strong agreement between the two codes. Both codes use Dougherty model collisions with $\nu_{ii} = 0.01\,v_{ti}/a$. This test corresponds to Fig. 5 of \citet{Gonzalez-Jerez2022}. {\color{black} As observed in \citet{Gonzalez-Jerez2022}, the discontinuity in the frequency is due to a change in mode structure for two branches of ITG. Both branches have very similar growth rates at $k_y \rho_i = 1.0$, resulting in a small but likely inconsequential disagreement between the codes in the frequency of the fastest-growing mode.}

\subsubsection{Nonlinear results}
Moving on to the nonlinear version of this test case, \cref{fig:nl_w7x} shows time traces of ion heat flux from \GX along with the time traces from stella and GENE, which were taken directly from the supplementary data made available by \citet{Gonzalez-Jerez2022} (and corresponding to Fig. 12 in that work{\color{black}; however note the factor of $\sqrt{2}$ difference in the definition of $v_{ti}$, and hence a factor of $2\sqrt{2}$ difference in $Q_{GB}$, in GX relative to the definitions used in stella and GENE}). Both stella and GENE used rather high velocity resolution for this case, with $N_{v_\parallel}=60$, $N_\mu=24$. For \GX we show a case with moderate velocity-space resolution, $( N_\ell, N_m)=(8,16)$, and a coarse resolution case with $( N_\ell, N_m)=(4,8)$. Both of these agree well with the higher-resolution cases from both stella and GENE, which indicates that the promising Laguerre-Hermite convergence observed in the tokamak benchmarks carries over to stellarators. Additionally, the wallclock time for the \GX cases was 90 minutes (on 4 GPUs) in the moderate resolution case and 28 minutes (on 1 GPU) in the coarse case.

\section{Velocity-space convergence and GPU performance}
\label{sec:convergence}

Part of the motivation for the Laguerre-Hermite pseudo-spectral approach is the ability to successfully run at low velocity-space resolution without uncontrolled approximations, since in the lowest-resolution limit the system corresponds to established gyrofluid models like the one of \citet{Beer1996}{\color{black}, albeit without Beer \& Hammett's collisionless closure schemes}.  
Another motivation for the pseudo-spectral approach is its fit for GPU computing, since the spectral algorithm relies heavily on fast transform methods that are well-optimized on GPUs, and the memory requirements of the algorithm are low enough to fit a problem onto one or a few GPUs. We also {\color{black} make exclusive use of} single-precision arithmetic, which is often sufficient for turbulence calculations \citep{Maurer2020}; there are no matrix inversions in our algorithm that would require higher precision{, \color{black} and we do not presently target simulations spanning both ion and electron scales that may stress the limits of single precision. The success of the benchmarks presented in the previous section is an additional indication that single-precision arithmetic is sufficient for a wide range of linear and nonlinear calculations.}

In this section we examine the velocity-space convergence for the nonlinear CBC calculations with \GX along with the performance of the calculations on one or several GPUs. Convergence and performance are complementary factors towards enabling first-principles transport calculations that are fast enough to be used for fusion reactor design {\color{black} along with wide-ranging parameter scans for physics discovery}. In this section we will present results detailing convergence and performance for the nonlinear CBC benchmark cases from \cref{sec:nl_cyclone}.

\begin{table}
    \centering
    \begin{tabular}{c c c c c c c}
    $ N_\ell$ & $ N_m$ & \qquad\quad$Q_i/Q_{GB}$ \qquad\qquad & wallclock (min) & time/step (s) & $\langle\Delta t\rangle$ & $N_\mathrm{gpu}$\\
\toprule
16 & 32	  & 7.8	$\pm$ 0.8	&58	&0.033 & 0.009 & 8 \\
\hline
$\star$\ \ 8\quad\quad &  16  & 7.3 $\pm$	0.9&	15&	    0.017 & 0.019 & 4\\
\hline
6	 &  12  & 7.7 $\pm$	0.8&	11	      &  0.017 & 0.026 & 2\\
\hline
$\star$\ \ 4\quad\quad  &  8   &7.2 $\pm$ 0.8	&5.5	&0.014 & 0.041 & 1\\
\hline
4 &  6   &7.0 $\pm$ 0.9	&3.5	&0.011 & 0.053 & 1\\
\hline
\hline
4	 &  4   &5.9 $\pm$ 0.7	&2.1	&0.0089 & 0.071 & 1\\
\hline
3	 &  8   &9.1 $\pm$ 0.8	&4.3	       & 0.011 & 0.042 & 1\\
\end{tabular}
    \caption{Velocity-space convergence and performance for the nonlinear CBC with Boltzmann electrons. Here $ N_m$ is the number of Hermite modes and $ N_\ell$ is the number of Laguerre modes. Other resolution parameters were $N_x=192,\ N_y=64,\ N_z=24$. Accurate ion heat flux calculations can be obtained with resolution as coarse as $( N_\ell, N_m)\geq(4,6)$ (above the double bar). Each simulation was run to $t = 1000 \,a/v_{ti}$, and we report the total wallclock time in minutes, the time per timestep in seconds, and the average timestep size $\langle\Delta t\rangle$ (normalized to $a/v_{ti}$), which changes with $ N_\ell$ and $ N_m$ due to linear stability constraints. The number of NVIDIA A100 GPUs used for each calculation is listed in the final column. The resolutions shown in \cref{fig:nl_cyclone} are marked with a $\star$.}
    \label{tab:nl_cyclone}
\end{table}

\begin{figure}
    \centering
    \includegraphics[width=\textwidth]{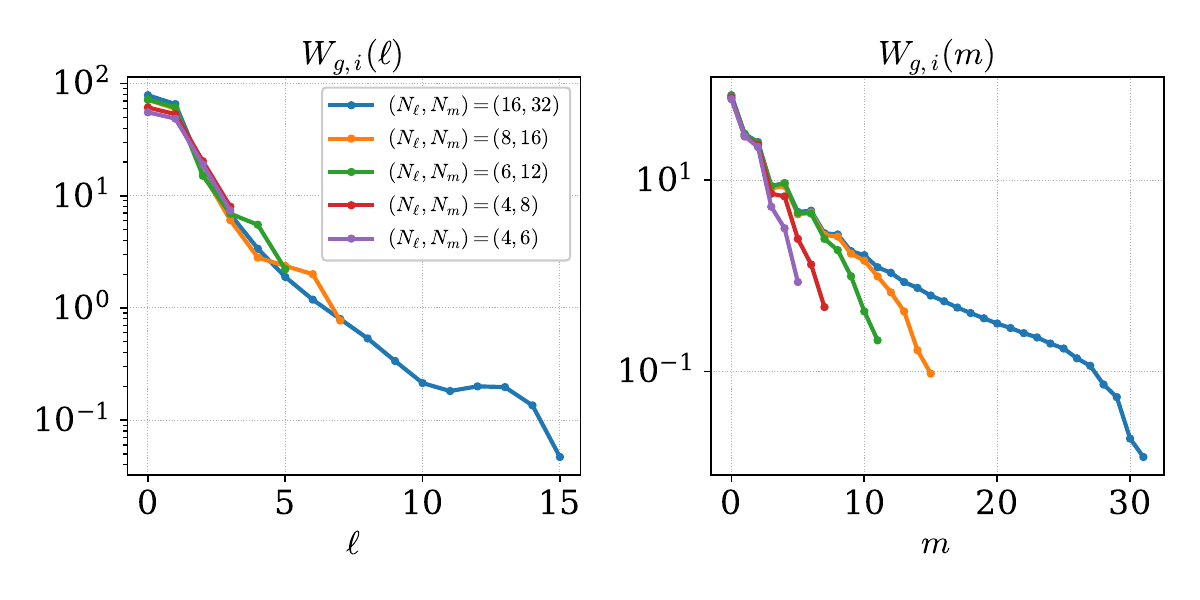}
    \caption{Laguerre (left) and Hermite (right) free energy spectra, $W_g$, for the nonlinear CBC with a Boltzmann electron response with varying velocity-space resolution. All the spectra are nearly identical at the scales that contribute dominantly to the heat flux (small $\ell$ and $m$), which is consistent with the excellent convergence observed in \cref{tab:nl_cyclone}.}
    \label{fig:nl_cyclone_lh}
\end{figure}

Starting with the case with Boltzmann electrons, \cref{tab:nl_cyclone} shows the results of several \GX calculations, with coarsening velocity resolution as we progress down the table. For this problem the convergence of the Laguerre-Hermite basis is quite remarkable, allowing accurate results with resolution as coarse as $( N_\ell, N_m) = (4,6)$. This is still a factor of four more moments than used in the \citet{Beer1996} six-moment gyrofluid model, but it is quite coarse relative to typical gyrokinetic calculations {\color{black} with $\mathcal{O}(10-100)$ velocity grid points}. Thus, the flexibility of the Laguerre-Hermite representation to smoothly interpolate between gyrofluid and gyrokinetic resolution enables the full nonlinear CBC calculation to be run accurately in less than 4 minutes on a single GPU, a remarkable result.

When we look at the Laguerre and Hermite spectra for these cases shown in \cref{fig:nl_cyclone_lh}, where $W_{g,s}(\ell,m) = \int |G_{\ell,m}^s|^2\, dx\, dy\, dz$ is the free energy in each moment, {\color{black} and $W_{g,s}(\ell) = \sum_m W_{g,s}(\ell, m)$ and $W_{g,s}(m) = \sum_\ell W_{g,s}(\ell, m)$}, we can see why the convergence is very good: even for the coarsest resolution, the spectra at the scales that contribute dominantly to the heat flux (small $\ell$ and $m$) are nearly identical. Note that hyper-collisions are being used in these calculations, and this is helping to enhance convergence. Without hyper-collisions (not shown), the results are only accurate with $( N_\ell, N_m) \geq (6,12)$. This means that hyper-collisions enable a five-fold reduction in simulation (wallclock) time for this case.

\begin{figure}
    \centering
    \includegraphics[width=\textwidth]{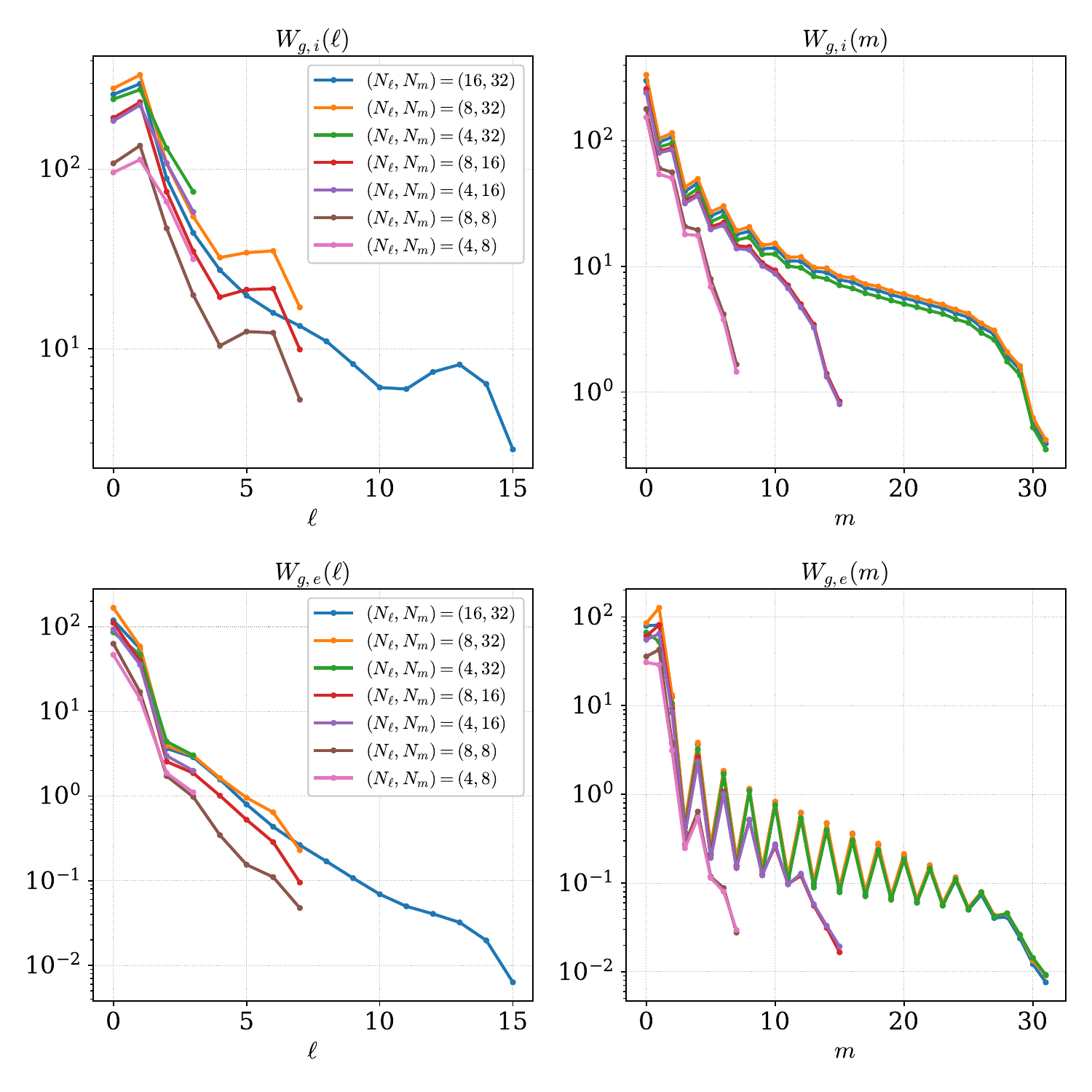}
    \caption{Laguerre (left) and Hermite (right) spectra for ions (top row) and electrons (bottom row) for the nonlinear CBC with kinetic electrons with varying velocity-space resolution.}
    \label{fig:nl_cyclone_kin_lh}
\end{figure}

We perform a similar convergence study for the CBC with kinetic electrons, as shown in \cref{tab:nl_cyclone_kin}. As above, $ N_\ell=4$ is sufficient for accurate heat flux predictions in both ion and electron channels. However, Hermite convergence is not as strong here, with reasonable accuracy {\color{black} (within 15\% of the highest resolution case)}  requiring $ N_m\gtrsim 16$. Examining the Laguerre and Hermite free energy spectra for these cases in \cref{fig:nl_cyclone_kin_lh}, we see more structure in the Hermite spectra, especially for the electrons. The oscillatory nature of the {\color{black} electron} spectrum, with higher amplitudes in even Hermite modes than odd, could be an indication that the toroidal drifts (which couple every other Hermite mode) are playing a strong role in the dynamics. {\color{black} This is consistent with the results of analytical and numerical convergence studies of the Laguerre-Hermite basis associated with toroidal drifts by \citet{Frei2022a,Frei2023}.}
Despite this structure, reasonably accurate results can be achieved in 159 minutes on 4 GPUs using $( N_\ell, N_m)=(4,16)$. This is quite good for a nonlinear, electromagnetic gyrokinetic ITG simulation with real-mass-ratio kinetic electrons. Additional details about multi-GPU scaling are presented in \cref{sec:scaling}.

\begin{table}
    \centering
    \begin{tabular}{c c c c c c c c}
    $ N_\ell$ & $ N_m$ & \qquad$Q_i/Q_{GB}$ \qquad\qquad & \quad$Q_e/Q_{GB}$ \qquad & wallclock (min) & time/step (s) & $\langle\Delta t\rangle$& $N_\mathrm{GPU}$\\
\toprule
$\star$\ \ 16\quad\quad	    &32 &   27.8    $\pm$   4.0 &	8.4	$\pm$	1.2	&	900&	0.068	& 0.00076 &	8 \\
\hline
8       &32	&   29.4    $\pm$   3.3	&	8.8	$\pm$	1.0	&	448	&	0.034		& 0.00076&   8 \\
\hline
4       &32	&   24.8    $\pm$   3.3	&	7.8	$\pm$	1.0	&	255	&	0.020	& 0.00077&	8 \\
\hline
8       &16	&   25.1    $\pm$   3.0	&	7.6	$\pm$	0.9	&	277	&	0.032	& 0.0012 &	4 \\
\hline
$\star$\ \ 4\quad\quad	    &16	&   23.6    $\pm$   3.2	&	7.3	$\pm$	0.9	&	159	&	0.019 & 0.0012	&	4\\
\hline
\hline
8       &8	&   18.6    $\pm$   2.1	&   6.4	$\pm$	0.7	&	159	&	0.03 & 0.0019	&	2 \\
\hline
4       &8	&   16.2    $\pm$   2.2	&	5.7	$\pm$	0.7	&	84	&	0.016 & 0.0019	&	2\\
\end{tabular}
    \caption{Velocity-space convergence and performance for the nonlinear CBC with kinetic electrons. Here $ N_m$ is the number of Hermite modes and $ N_\ell$ is the number of Laguerre modes. Other resolution parameters were $N_x=192,\ N_y=64,\ N_z=24$. Reasonably accurate heat flux calculations {\color{black} (within 15\% of the highest resolution case)} can be obtained with resolution as coarse as$( N_\ell, N_m)\geq(4,16)$ (above the double bar). Each simulation was run to $t = 600 \,a/v_{ti}$, and we report the total wallclock time in minutes, the time per timestep in seconds, and the average timestep size $\langle\Delta t\rangle$ (normalized to $a/v_{ti}$). The number of NVIDIA A100 GPUs used for each calculation is also shown. The resolutions used for the GX calculations shown in \cref{fig:nl_cyclone_kin} are marked with a $\star$.}
    \label{tab:nl_cyclone_kin}
\end{table}

Note that using more Hermite modes not only adds to the size of the problem, but also makes the Courant-Friedrichs-Levy (CFL) constraint on the timestep more restrictive because the maximum parallel velocity $v_{\parallel\mathrm{max}}$ on the Hermite collocation grid increases with the number of Hermite modes. In future work we will try to alleviate this issue by constraining the maximum velocity on the collocation grid. However, a commonly-used method of scaling the argument of the Hermite polynomials to reduce the extents of the collocation grid is ill-suited to our problem, since the resulting projection of a Maxwellian onto the scaled basis often diverges \citep{Fok2001}. Instead, an approach similar to that used by \citet{Candy2016} to bound the energy grid in CGYRO could be used here.

\subsection{GPU scaling studies} \label{sec:scaling}
\begin{figure}
    \centering
    \includegraphics[height=2in]{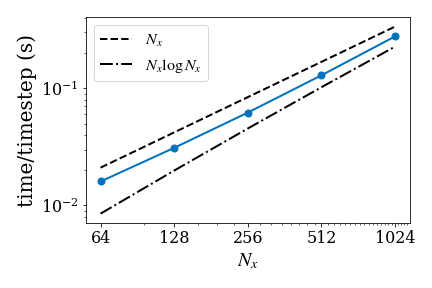}
    \caption{Scaling of time per timestep (in seconds) vs the number of radial grid points, $N_x$, for nonlinear CBC with Boltzmann electrons on a single GPU. Ideal linear scaling is shown with the dashed line, while the dot-dashed line shows $N_x \log N_x$ scaling (the expected scaling when FFTs dominate the algorithm).}
    \label{fig:nx_scaling}
\end{figure}

For single-GPU calculations, the GPU threading architecture effectively enables dynamic parallelism over the entire phase space. In \cref{fig:nx_scaling}, we show that for a nonlinear CBC calculation with Boltzmann electrons, the algorithm scales roughly linearly with the number of radial grid points, $N_x$. This is better than the $\mathcal{O}(N\log N)$ cost that would be expected when the fast Fourier transform (FFT) in the pseudo-spectral algorithm dominates the cost. This is also better scaling than other gyrokinetic algorithms that require $\mathcal{O}(N^2)$ operations due to matrix inversion.

When considering multi-GPU parallelization, the cost of inter-GPU memory transfers can be quite large compared to floating point operations. Thus it is important to design the scheme to minimize memory transfers and take advantage of the capability of modern GPUs to overlap communication with computation. Thus in \GX we have chosen a targeted multi-GPU parallelization scheme that currently parallelizes only species and Hermite modes across GPUs. In this scheme, communication is required to compute the total charge density and currents in the field equations, which takes the form of an all-reduce operation on a configuration-space-sized ($N_{k_x}\times N_{k_y}\times N_z$) object. Additionally, since the equation for the Hermite mode $m$ couples to modes $m-2$, $m-1$, $m+1$ and $m+2$, we use ``halo'' (or ``ghost'') modes on each GPU, similar to standard parallelization schemes for finite differencing. Halo exchange is thus the other dominant form of communication, requiring inter-GPU transfers of objects of size $N_{k_x}\times N_{k_y}\times N_z\times N_\ell\times 2$. We overlap the halo exchange with computation of most of the nonlinear terms (halo modes are only required for the nonlinear terms involving $A_\parallel$, and even for these the terms on the interior modes can be computed before the halo transfers have been completed). 

\begin{figure}
    \centering
    \includegraphics[height=2in]{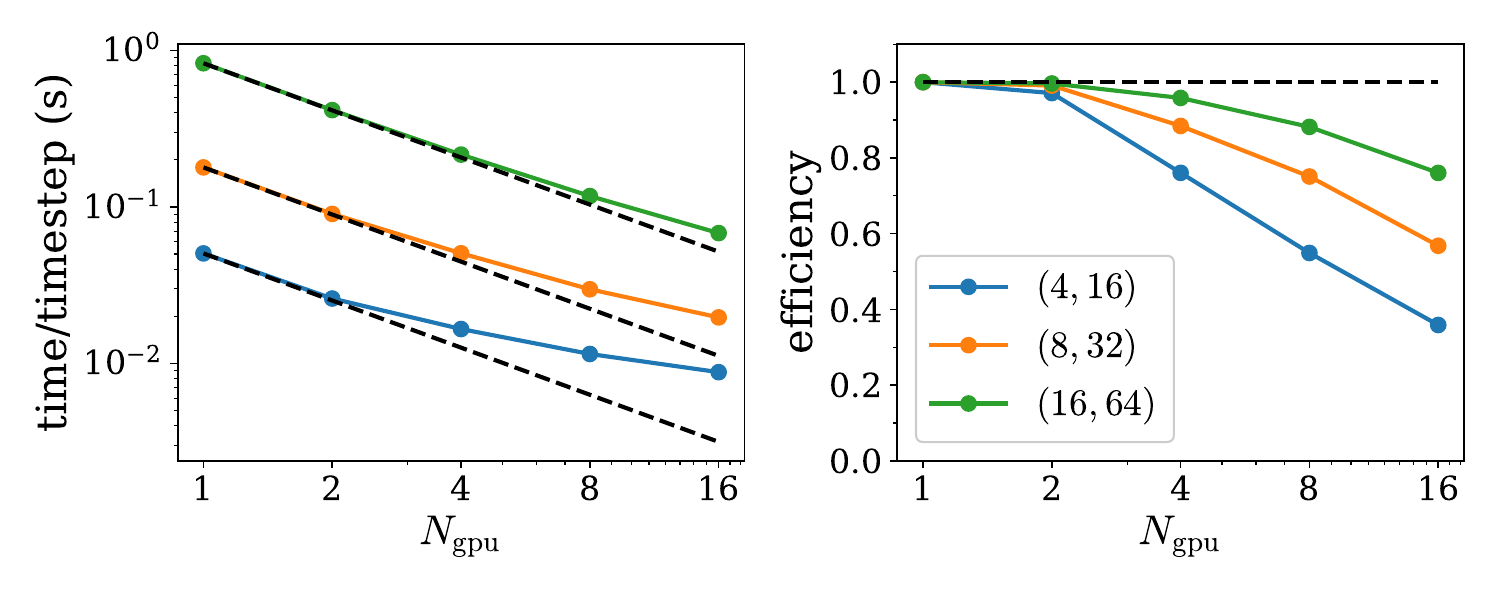}
    \caption{Strong scaling study for the nonlinear CBC with kinetic electrons showing the time per timestep (a) and the scaling efficiency (b) as a function of number of GPUs used, with fixed resolution for each curve. In each case the ideal scaling is shown with a dashed black line. {\color{black} The Laguerre-Hermite resolution parameters are listed in the legend as $(N_\ell, N_m)$. The other resolution parameters are: $N_x=192,\ N_y=64,\ N_z=24,\ N_\mathrm{species}=2$.}
    }
    \label{fig:strong_scaling}
\end{figure}

In \cref{fig:strong_scaling} we show a strong scaling study for nonlinear CBC cases with {\color{black} two} kinetic species. In each case we fix resolution (shown in the legend) and plot the time per timestep (a) and scaling efficiency (b) as we increase the number of GPUs used for the calculation. These calculations have been performed on Perlmutter at NERSC, where each node contains four A100 GPUs. The three cases have the same configuration space resolution and number of kinetic species as the cases in \cref{tab:nl_cyclone_kin}. {\color{black} For reference, the (16,64) case running on a single GPU consumes about 90\% of the available GPU memory on a 40GB A100 GPU. In each case, the scaling efficiency is above 75\% parallelizing across 4 GPUs. The efficiency also improves as the problem size increases, which is expected because there is more compute time relative to communication time.}

\begin{figure}
    \centering
    \includegraphics[height=2in]{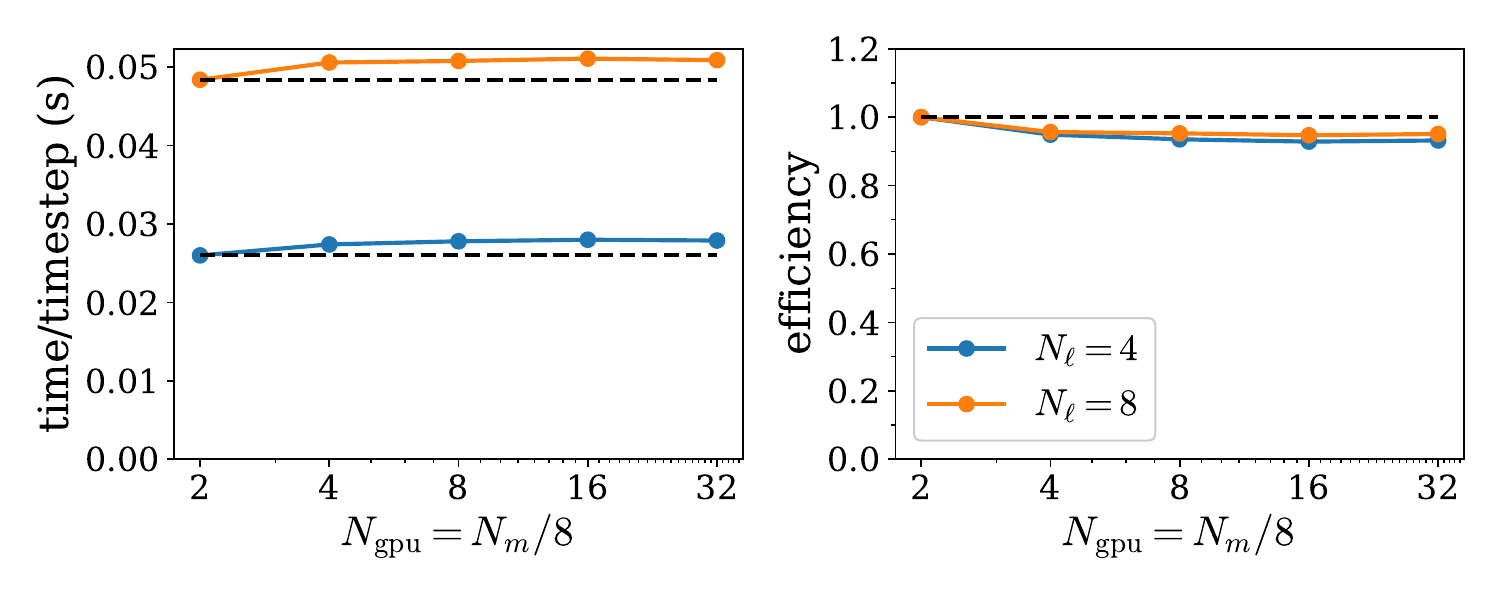}
    \caption{Weak scaling study for the nonlinear CBC with kinetic electrons showing the time per timestep (a) and the scaling efficiency (b) as a function of number of GPUs used, with the number of Hermite modes $ N_m$ scaling with the number of GPUs as $ N_m=8N_\mathrm{gpu}$. We show results for both $ N_\ell=4$ (blue) and $ N_\ell=8$ (yellow). The ideal scaling is shown with dashed lines. The other resolution parameters are: $N_x=192,\ N_y=64,\ N_z={\color{black} 24},\ N_\mathrm{species}=2$.}
    \label{fig:weak_scaling}
\end{figure}

We also show in \cref{fig:weak_scaling} weak scaling studies for the nonlinear CBC with two kinetic species and the same configuration space resolution as the cases in \cref{tab:nl_cyclone_kin}. Here, we increase the number of Hermite modes proportionally to the number of GPUs used, so that $ N_m=8 N_\mathrm{gpu}$ in each case. After parallelizing first over the two kinetic species, this results in 16 Hermite modes per GPU in all the calculations shown. We show the weak scaling for both $ N_\ell=4$ (blue) and $ N_\ell=8$ ({\color{black} yellow}). The weak scaling efficiency, shown in (b), is ideal {\color{black} in both cases up to 32 GPUs ($N_m = 256$), despite the fact that using more than 4 GPUs requires multiple nodes on Perlmutter, and communication between nodes can be slower than communication within a node due to the details of the interconnect hardware.  
}

\section{Conclusion \& future opportunities} \label{sec:conclusion}

In this work we have presented \GX, a GPU-native gyrokinetic code focused on tokamak and stellarator design at reactor scales. We have described the numerical algorithms that we use to solve the electromagnetic $\delta\!f$ gyrokinetic system, in particular the discretization scheme that is pseudo-spectral in the entire five-dimensional phase space, leveraging the Laguerre-Hermite velocity-space formulation developed by \citet{Mandell2018}. We have shown several linear and nonlinear benchmarks against established flux-tube gyrokinetic codes in both tokamak and stellarator geometry, verifying that \GX correctly simulates gyrokinetic turbulence in the small $\rho_*$ limit. The combination of GPU acceleration and favorable convergence properties of the Laguerre-Hermite velocity-space basis for nonlinear problems enables useful turbulence simulations in minutes.  

Additional work will also focus on improving the efficiency of simulations that include kinetic electrons. At present, the need to resolve the fast parallel electron motion results in an increase in simulation cost by a factor of roughly $2\sqrt{m_i/m_e}\approx 120$ compared to simulations with Boltzmann electrons (the factor of 2 is from the additional kinetic species, and the square root of the mass ratio comes from $v_{te}/v_{ti}$). We have also shown that kinetic electron cases may require more Hermite resolution, especially if resolving the trapped-passing boundary is essential. Exploration of different choices of velocity-space coordinates for the electrons or improvements to the basis functions used (for example, a method to bound the Hermite collocation points to constrain $v_{\parallel\mathrm{max}}$, as in \citet{Candy2016}, which could alleviate the CFL timestep restriction) could increase efficiency. In addition, the reduced electron models of \citet{Beer1996, Snyder2001, Abel2013a} provide insight into how to eliminate the fast timescales associated with parallel electron motion. A combination of these approaches with implicit timestepping schemes could enable kinetic electron simulations with nearly the same efficiency as the Boltzmann electron simulations we have presented. {\color{black} Additionally, we are investigating machine-learning methods for sub-grid models and closures \citep{barbour2021,barbour2022}, as well as methods for dynamically adapting the number of modes in the system.}

An essential target for \GX is transport time-scale macro-scale profile evolution via coupling to a transport solver like Trinity \citep{Barnes2009, Qian2022} {\color{black} or a steady-state profile prediction solver like TGYRO \citep{Candy2009} or PORTALS \citep{Rodriguez-Fernandez2022}}. This will be described in a forthcoming paper. In this case, several \GX flux-tube calculations are run in parallel at various radii, and the resulting turbulent fluxes are used to advance the transport equations for the equilibrium profiles on transport time-scales. The efficiency of \GX makes these simulations quite tractable in comparison to previous efforts. Further, since transport in a fusion reactor is usually quite stiff, small changes in profile gradients result in large changes in transport fluxes; conversely, this means that errors of order 10\% are quite tolerable since they will result in relatively negligible differences in the final equilibrium profiles. Thus we can also leverage the flexibility of \GX to run at quite low velocity resolution with only 10-20\% errors in the turbulent fluxes, as shown in the convergence studies.

Finally, since \GX is specialized for reactor design, using \GX as a turbulence model in the inner loop of optimization frameworks is of great interest. \citet{Kim2024} have leveraged the DESC \citep{dudt2022} optimization framework to demonstrate that stellarators can be optimized for turbulence using nonlinear \GX calculations in the optimization loop. Additionally, coupling of \GX to the SIMSOPT stellarator optimization framework \citep{Landreman2021} is in progress. Following the work of \citet{Highcock2018}, optimization of core equilibrium profiles (and hence fusion power) is also tractable with a fast multi-scale transport model composed of \GX coupled to a transport solver.

\section{Acknowledgements}

The authors thank G. Hammett, M. Landreman, M. Zarnstorff, B. Buck, N. Barbour, J. Parisi, M. Barnes, and F. Parra for helpful discussions and encouragement.
Research support came from the U.S. Department of Energy (DOE) and the U.S. National Science Foundation (NSF): N.R.M. was supported by the DOE Fusion Energy Sciences Postdoctoral Research Program
administered by the Oak Ridge Institute for Science and Education (ORISE) for the DOE via Oak Ridge
Associated Universities (ORAU) under DOE contract number DE-SC0014664 and by the Laboratory Directed Research and Development Program of the Princeton Plasma Physics Laboratory under U.S. Department of Energy contract number DE-AC02-09CH11466; W. D., R. G., and P. K. were supported by DOE via the Scientific Discovery Through
Advanced Computing Program under award number DESC0018429; P.K is also supported by the DOE CSGF Program under award number DE-SC0024386; T. Q. was supported by NSF GRFP Grant No. DGE-2039656.  
Computations were performed on the Stellar cluster at Princeton University/PPPL and the Perlmutter cluster at NERSC. 
All opinions expressed in this paper are the authors'
and do not necessarily reflect the policies and views of DOE, NSF, ORAU, or ORISE.

\section{Declaration of Interests} The authors report no conflict of interest.

 \appendix

\section{Normalization} \label{app:norm}
We choose a set of normalizations, described in Table~\ref{tab:Table-fundamental-norm}, for the fundamental quantities.  
\begin{table}
\centering
\begin{tabular}{c|c|c|c|c|c}
 $\rm{length}$ (m) & $\rm{magnetic\, field}$ (T) & $\rm{temperature}$ (K) & $\rm{mass}$ (kg) & $\rm{number} \, \rm{density}$ (m$^{-3}$) & $\rm{charge}$ (C) \\[3pt]
 $a_\mathrm{N}$  & $B_\mathrm{N}$  & $T_{\rm{ref}}$ & $M_{\rm{ref}}$ & $n_{\rm{0}}$ & $q_{\rm{ref}}$\\
\end{tabular}
\caption{Fundamental normalizing quantities}
\label{tab:Table-fundamental-norm}
\end{table}
{\color{black} Note that the specific definition of the normalizing quantities is arbitrary; for example, the normalizing length could be chosen to be the minor radius ($a_N = a$), the major radius ($a_N = R$), or some other length.} 
Using the fundamental normalizations, we can define the reference thermal speed $v_{\rm{th, ref}} = \sqrt{T_{\rm{ref}}/M_{\rm{ref}}}$, the reference cyclotron frequency $\Omega_{\rm{ref}} = (q_{\rm{ref}} B_{\rm{N}})/(M_{\rm{ref}}c)$ and the reference gyroradius $\rho_{\rm{ref}}  \equiv v_{\rm{th, ref}}/\Omega_{\rm ref}$, with $c$ the speed of light. The ratio of the reference gyroradius $\rho_{\rm{ref}}$ and the normalizing length $a_{\rm{N}}$ is defined as
\begin{equation}
    \rho_{*} \equiv \frac{\rho_{\rm{ref}}}{a_{\rm{N}}},
\end{equation}
which is the fundamental small parameter in the gyrokinetic model. Using Table~\ref{tab:Table-fundamental-norm}, we can also normalize all the independent variables and operators as shown in Table~\ref{tab:Table-independent-norm} and all the dependent variables as shown in Table~\ref{tab:Table-dependent-norm}.
\begin{table}
\centering
\begin{tabular}{c|c|c|c|c|c}
 $k_x, k_y, x, y$ & $\boldsymbol{\nabla}, z$ & $v_\parallel$ & $\mu$ & $\omega, t$  & $\psi$ \\[3pt]
 $\rho_{\rm{ref}}$  & $a_{\rm{N}}$ & $v_{\rm{th},s}$ & $v_{\rm{th},s}^2{\color{black}/B_\mathrm{N}}$  & $v_{\rm{th, ref}}/a_{\rm{N}}$ & $a_{\rm{N}}^2B_{\rm{N}}$\\
\end{tabular}
\caption{Normalizing the quantities for the independent variables}
\label{tab:Table-independent-norm}
\end{table}
\begin{table}
\centering
\begin{tabular}{c|c|c|c}
 $h$ & $\Phi$ & $A_{\parallel}$ & $\delta\!B_{\parallel}$  \\[3pt]
 $\rho_* n_0/v^3_{\rm{th, ref}}$  & $\rho_{*}T_{\rm{ref}}/q_{\rm{ref}}$  & $\rho_{*}(\rho_\mathrm{ref} B_{\rm{N}})$ & $\rho_{*} B_{\rm{N}}$\\
\end{tabular}
\caption{Normalizing quantities for dependent variables}
\label{tab:Table-dependent-norm}
\end{table}
Using the normalizations given in Tables~\ref{tab:Table-independent-norm} and~\ref{tab:Table-dependent-norm}, we can normalize the gyrokinetic model. Note that the term $\rho_{*}$ cancels out of all the equations once they are normalized. We also define the following non-dimensional species-specific parameters that appear throughout the normalized equations: $\tau_s = T_s/T_\mathrm{ref}$, $m_s = M_s/M_\mathrm{ref}$, $v_{ts} = v_{\mathrm{th},s}/v_\mathrm{th,ref}$, $Z_s = q_s/q_\mathrm{ref}$, and $\rho_s = (v_{\mathrm{th},s}/\Omega_s)/\rho_\mathrm{ref}$.

{\color{black}
\section{Derivation of hypercollision operator} \label{app:hyper}
Consider the simplified system where the distribution function $g$ evolves due to only parallel free streaming. Fourier transforming in the parallel direction, a mode with parallel wavenumber $k$ evolves via
\begin{equation}
    \pderiv{g}{t} + i k v_t v_\parallel g = 0.
\end{equation}
The solution is 
\begin{equation}
    g(k,v,t) \propto e^{-i k v_t v_\parallel t}
\end{equation}
Free streaming thus causes energy transfer to smaller scales in velocity space; this is the parallel phase mixing behavior that produces Landau damping, as is well known. In the Hermite representation results this results in transfer of energy to higher $m$, which can be seen by taking Hermite moments of the solution, giving
\begin{equation}
    g_m(k,t) \propto t^m e^{- k^2 v_t^2 t^2/2}.
\end{equation}
Thus a pulse propagating in $m$ reaches $m(t) = k^2 v_t^2 t^2$ at time $t$. By differentiating in time we can obtain the advection ``velocity'' in $m$-space:
\begin{equation}
    \frac{d m}{dt} = 2 k^2 v_t^2 t = 2 k v_t \sqrt{m}. \label{madv}
\end{equation}
Now consider the advection equation for the mode energy, $E(m) = |g_m|^2/2$,
\begin{equation}
    \pderiv{E}{t} + \pderiv{}{m}\left[\frac{dm}{dt} E\right] = S + D
\end{equation}
The source $S$ only acts at low $m$. For dissipation, we will assume a hypercollision operator of the form
\begin{equation}
    \left(\pderiv{g_m}{t} \right)_\mathrm{hyp} = - \nu_\mathrm{hyp} m^p g_m,
\end{equation}
which results in
\begin{equation}
    D = - 2 \nu_\mathrm{hyp} m^p E.
\end{equation}
In steady state ($\partial/\partial t \rightarrow 0$) at large enough $m$ to neglect the source, we have
\begin{equation}
    \pderiv{}{m}\left[\frac{dm}{dt} E\right] = - 2 \nu_\mathrm{hyp} m^p E.
\end{equation}
Inserting the advection velocity from \cref{madv}, we obtain
\begin{equation}
     \pderiv{}{m}\left[2 k v_t \sqrt{m} E\right] = - 2 \nu_\mathrm{hyp} m^p E.
\end{equation}
The solution of this differential equation is of the form
\begin{equation}
    E(m) = \frac{C}{\sqrt{m}} \exp\left({-\frac{\nu_\mathrm{hyp} m^{p+1/2}}{(p+1/2)k v_t}}\right)
\end{equation}
Note that in the limit of no dissipation ($\nu_\mathrm{hyp}\rightarrow 0$), we should have a constant $m-$flux $\Gamma$, which restricts the constant of integration to $C=\Gamma/(2k v_t)$ and makes the full solution
\begin{equation}
    E(m) = \frac{\Gamma}{2k v_t\sqrt{m}} \exp\left({-\frac{\nu_\mathrm{hyp} m^{p+1/2}}{(p+1/2)k v_t}}\right).
\end{equation}
Now we would like to choose $\nu_\mathrm{hyp}$ so that the energy is damped to some fraction $f$ by the time the energy reaches the grid scale at $m=m_\mathrm{max}$. From this we obtain
\begin{equation}
    \nu_\mathrm{hyp} = \ln(f^{-1}) \frac{p+1/2}{m_\mathrm{max}^{p+1/2}}|k|v_t
\end{equation}
Taking $f=0.1$ results in the operator presented in \cref{sec:hyper}. Note that undamped energy will be reflected and damped again as it travels from high $m$ to low $m$, resulting in only $f^2=0.01$ of the initial energy returning to low $m$.
}

{\color{black}
\section{Convergence study for parallel boundary damping filter} \label{app:damp}

In \cref{sec:bcs} we defined the parallel boundary damping filter using two adjustable parameters: the width of the damping region $z_\mathrm{width}$, and the scaling factor $A$. Here we perform convergence studies to verify that we have chosen reasonable values of these parameters. We will use the kinetic electron CBC for these studies.

We first perform a scan of the scaling factor $A$. \cref{fig:damp_amp_scan} shows the ion and electron heat flux as a function of $A\Delta t$, with $\Delta t$ the timestep size. For this scan we have used $z_\mathrm{width} = L_z/8$. We can see that the heat fluxes are converged when $A\Delta t \gtrsim 0.05$. We next perform a scan of the damping region width $z_\mathrm{width}$, as shown in \cref{fig:damp_width_scan}. For this scan we use $A\Delta t = 0.1$. By varying $z_\mathrm{width}/L_z$ for several values of $N_z$, we find that $z_\mathrm{width} = L_z/8$ produces accurate results for each choice of $N_z$.

\begin{figure}
    \centering
    \includegraphics[width=.6\textwidth]{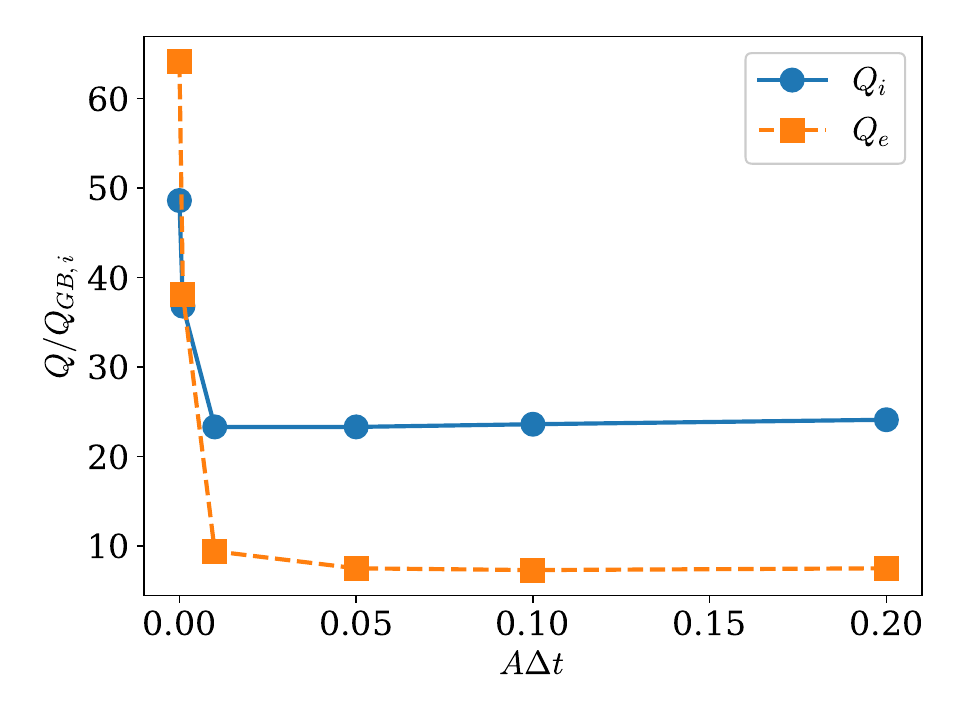}
    \caption{Time-averaged gyro-Bohm-normalized ion (solid blue circles) and electron (dashed yellow squares) heat fluxes as a function of $A\Delta t$, with $A$ the scaling factor for the damping filter and $\Delta t$ the timestep size.}
    \label{fig:damp_amp_scan}
\end{figure}

\begin{figure}
    \centering
    \includegraphics[width=.6\textwidth]{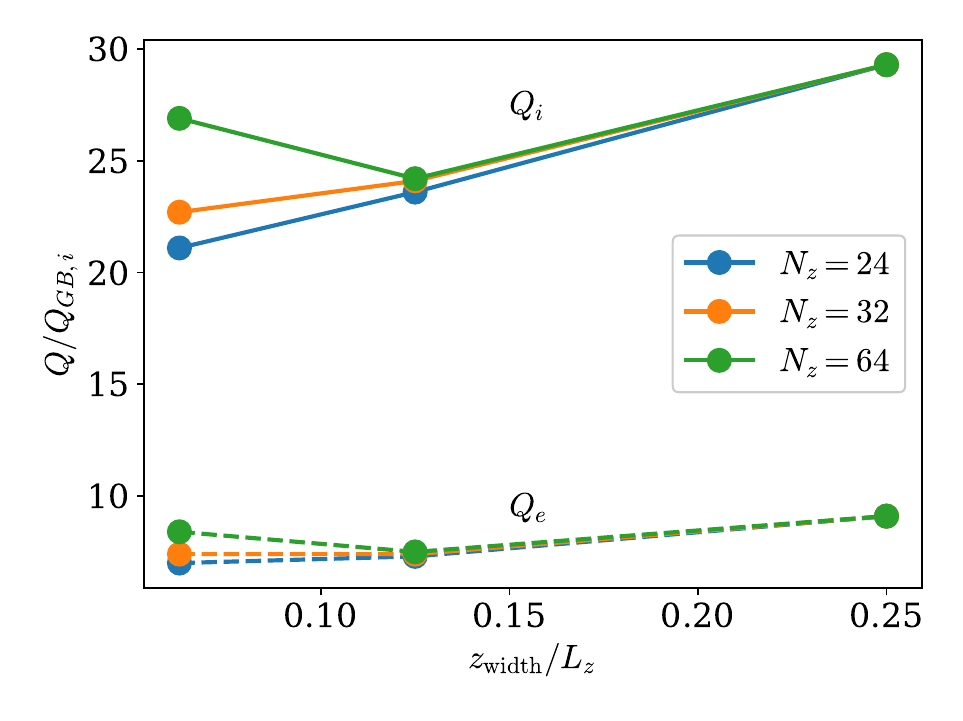}
    \caption{Time-averaged gyro-Bohm-normalized ion (solid) and electron (dashed) heat fluxes as a function of $z_\mathrm{width}/L_z$, with $z_\mathrm{width}$ the width of the damping region and $L_z$ the parallel length of the flux tube. Colors indicate different values of parallel resolution $N_z$.}
    \label{fig:damp_width_scan}
\end{figure}
}

\section{Pseudo-spectral evaluation of nonlinear terms} \label{app:nl}

The nonlinear terms are evaluated pseudo-spectrally in $(x,y,z,\mu B, m)$ space to avoid convolutions in both Fourier and Laguerre coefficients as \citep{Mandell2018},
\begin{align}
    \mathcal{N}_{\ell,m}^s &= \mathcal{L}_\ell \mathcal{H}_m \mathcal{F}_{\bf k_\perp} [\gyavgR{\vEM}\cdot\nabla h_s] = \mathcal{L}_\ell \mathcal{H}_m \mathcal{F}_{\bf k_\perp} [\gyavgR{\vEM}\cdot\nabla g_s] \notag \\
    &= \mathcal{L}_\ell \mathcal{F}_{\bf k_\perp}\left(\mathcal{F}_{\bf k_\perp}^{-1}[i k_y J_{0s}\Phi] \mathcal{F}_{\bf k_\perp}^{-1}[i k_x \sum_{\ell'} \psi^{\ell'} G^s_{\ell',m}] - \mathcal{F}_{\bf k_\perp}^{-1}[i k_x J_{0s}\Phi] \mathcal{F}_{\bf k_\perp}^{-1}[i k_y\sum_{\ell'} \psi^{\ell'} G^s_{\ell',m}] \right) \notag \\
    &\ - v_{ts}\sqrt{m+1}\mathcal{L}_\ell \mathcal{F}_{\bf k_\perp}\left(\mathcal{F}_{\bf k_\perp}^{-1}[i k_y J_{0s} A_\parallel] \mathcal{F}_{\bf k_\perp}^{-1}[i k_x \sum_{\ell'} \psi^{\ell'} G^s_{\ell',m+1}]\right.
    \notag \\ &\qquad\qquad\qquad\qquad\qquad\qquad
    \left.- \mathcal{F}_{\bf k_\perp}^{-1}[i k_x J_{0s} A_\parallel] \mathcal{F}_{\bf k_\perp}^{-1}[i k_y\sum_{\ell'} \psi^{\ell'} G^s_{\ell',m+1}] \right) \notag \\
    &\ - v_{ts}\sqrt{m}\mathcal{L}_\ell \mathcal{F}_{\bf k_\perp}\left(\mathcal{F}_{\bf k_\perp}^{-1}[i k_y J_{0s} A_\parallel] \mathcal{F}_{\bf k_\perp}^{-1}[i k_x \sum_{\ell'} \psi^{\ell'}  G^s_{\ell',m-1}]\right.
    \notag \\ &\qquad\qquad\qquad\qquad\qquad\qquad
    \left. - \mathcal{F}_{\bf k_\perp}^{-1}[i k_x J_{0s} A_\parallel] \mathcal{F}_{\bf k_\perp}^{-1}[i k_y\sum_{\ell'} \psi^{\ell'}  G^s_{\ell',m-1}] \right)\notag \\
    &\ +\frac{\tau_s}{Z_s}\mathcal{L}_\ell \mathcal{F}_{\bf k_\perp}\left(\mathcal{F}_{\bf k_\perp}^{-1}[i k_y 2\mu B \frac{J_{1s}}{\alpha_s}\delta\! B_\parallel] \mathcal{F}_{\bf k_\perp}^{-1}[i k_x \sum_{\ell'} \psi^{\ell'}  G^s_{\ell',m}] \right.
    \notag \\ &\qquad\qquad\qquad\qquad\qquad\qquad
    \left.- \mathcal{F}_{\bf k_\perp}^{-1}[i k_x 2\mu B \frac{J_{1s}}{\alpha_s}\delta\! B_\parallel] \mathcal{F}_{\bf k_\perp}^{-1}[i k_y \sum_{\ell'} \psi^{\ell'}  G^s_{\ell',m}] \right),
\end{align}
This requires zero-padding of the arrays in the perpendicular and Laguerre dimensions to avoid aliasing, as described in Appendix D of \cite{Mandell2018}. {\color{black} Fourier transforms are evaluated using the cuFFT library. The Laguerre transforms are expressed as matrix multiplications and implemented using the cuBLAS library.}

\section{Geometry details}
\label{app:geometry-appendix1}
In this appendix, we will describe how the geometry-related coefficients are calculated in the \GX code. First, we define the different coordinate systems and relevant identities and briefly outline the process of calculating the geometry-related coefficients. Next, we list all the geometric quantities used by \GX in the field-line following coordinate system. Finally, we describe the process of obtaining an equal-arc, equispaced coordinate from a general coordinate $z$. 

Consider a magnetic coordinate system $(\psi, \phi, \theta)$, where $\psi$ is a flux surface label, $\phi$ is the cylindrical toroidal angle, and $\theta$ is a generalized poloidal angle. A general 3D equilibrium is a union of nested flux surfaces labeled with the flux surface label $\psi$, which is usually produced by an equilibrium code in the cylindrical coordinate $(R, \phi, Z)$. The first step is to calculate the Jacobian from the cylindrical to the curvilinear coordinates $(\psi, \phi, \theta)$. 
\begin{equation}
    \mathcal{J}  = \frac{1}{(\boldsymbol{\nabla}\psi \times \boldsymbol{\nabla}\phi)\cdot \boldsymbol{\nabla}\theta} = \left(\frac{\partial \boldsymbol{R}}{\partial \psi} \times \frac{\partial \boldsymbol{R}}{\partial \phi}\right)\cdot\frac{\partial \boldsymbol{R}}{\partial \theta}
    \label{eqn:Jacobian-appendix}
\end{equation}
where the last equality in~\eqref{eqn:Jacobian-appendix} is a consequence of the duality relation~\citep{Dhaeseleer1991}
\begin{equation}
    \boldsymbol{\nabla}X_i \cdot \frac{\partial \boldsymbol{R}}{\partial X_j} = \delta_{ij},
    \label{eqn:Duality-relation}
\end{equation}
where $X_i$ and $X_{j}$ are coordinates. Using the Jacobian and the duality relation we calculate the gradients $\boldsymbol{\nabla} \psi, \boldsymbol{\nabla} \phi, \boldsymbol{\nabla} \theta$ from the gradients $\boldsymbol{\nabla} R,  \boldsymbol{\nabla} \phi,  \boldsymbol{\nabla} Z$. For example, we calculate $\boldsymbol{\nabla} \psi$ using the following relation
\begin{equation}
    \boldsymbol{\nabla} \psi = \frac{\left(\frac{\partial \boldsymbol{R}}{\partial \theta} \times \frac{\partial \boldsymbol{R}}{\partial \phi}\right)}{\left(\frac{\partial \boldsymbol{R}}{\partial \psi} \times \frac{\partial \boldsymbol{R}}{\partial \phi}\right)\cdot\frac{\partial \boldsymbol{R}}{\partial \theta}}
\end{equation}
where the vector $\boldsymbol{R}=(R, \phi, Z)$ is output from a typical equilibrium solver. Upon calculating the gradients of the coordinates $(\psi, \theta, \phi)$, we can easily obtain the gradients of the field-line-following coordinate $(x, y, z)$. In addition to that, an equilibrium code also outputs scalar-valued physical quantities like the plasma pressure $p(\psi)$, and the safety factor $q(\psi)$ and vector-valued physical quantities like $B, \boldsymbol{B}\cdot\boldsymbol{\nabla}\theta, \boldsymbol{B}\cdot\boldsymbol{\nabla}\phi$. These quantities are then used to compute all the geometric coefficients defined below:
\begin{equation}
    \texttt{bmag} = \frac{B}{B_{\rm{N}}}
\end{equation}

\begin{equation}
    \texttt{gradpar} = \boldsymbol{b}\cdot\bold{\nabla} z
\end{equation}

\begin{equation}
    \texttt{gds21} = \hat{s}(\boldsymbol{\nabla}x\cdot\boldsymbol{\nabla}y), \quad \hat{s} \equiv \frac{x}{q}\frac{dq}{dx}
\end{equation}

\begin{equation}
    \texttt{gds22} = \hat{s}^2\lvert\boldsymbol{\nabla}x\rvert^2
\end{equation}

\begin{equation}
    \texttt{gds2} = \lvert\boldsymbol{\nabla}y\rvert^2
\end{equation}

\begin{equation}
    \texttt{gbdrift} = \frac{2}{\texttt{bmag}^2}(\boldsymbol{b} \times \boldsymbol{\nabla}B)\cdot \boldsymbol{\nabla}y
\end{equation}

\begin{equation}
    \texttt{cvdrift} =  \frac{8\pi}{\texttt{bmag}^2}  (\boldsymbol{b} \times \boldsymbol{\nabla}y)\cdot \boldsymbol{\nabla}p + \texttt{gbdrift} 
    \label{eqn:cvdrift}
\end{equation}

\begin{equation}
    \texttt{cvdrift0} = \texttt{gbdrift0} =  \frac{2\hat{s}}{\texttt{bmag}^2}(\boldsymbol{b} \times \boldsymbol{\nabla}B)\cdot \boldsymbol{\nabla}x
\end{equation}
{\color{black} In the local flux-tube limit 
 we evaluate all of these quantities at $x=x_0$, with $x_0$ denoting the flux surface of interest. In non-axisymmetric configurations the coefficients also in general depend on the field line label $y$, so we also evaluate the coefficients at $y=y_0$ to select a particular field line of interest. This ensures that the geometric coefficients are only functions of the parallel coordinate $z$.} These quantities form a complete set of geometry-related coefficients needed for a \GX simulation. These quantities also coincide with the geometry definitions in GS2.
\subsection{Calculating an equispaced, equal-arc z} \label{eqarc}
Generally speaking, $\texttt{gradpar}$ is a function of $z$. However, it is often convenient to transform to a $z$-grid where $\texttt{gradpar}$ is a constant. This is strictly necessary in \GX to allow Fourier spectral treatment of the parallel streaming term in the gyrokinetic equation. To do this, we seek a grid $\hat{z}$ such that
\begin{equation}
    \boldsymbol{b}\cdot \boldsymbol{\nabla} \hat{z} = C,
\end{equation}
where $C$ is a constant. A differential element in the field-line following coordinates $(x, y, z)$ can be written as
\begin{equation}
    d\boldsymbol{\ell} = \pderiv{\vec{R}}{x}dx + \pderiv{\vec{R}}{y}dy + \pderiv{\vec{R}}{z}dz.
\end{equation}
On a given field line, $dx = dy = 0$. Using the duality relation~\eqref{eqn:Duality-relation}, we can write
\begin{equation}
    d\boldsymbol{\ell} = \frac{\boldsymbol{\nabla}x\times \boldsymbol{\nabla}y}{(\boldsymbol{\nabla}x\times \boldsymbol{\nabla}y)\cdot\boldsymbol{\nabla}z} dz.
    \label{eqn:Duality-field-line-following}
\end{equation}
For a differential line element along the field line, we dot~\eqref{eqn:Duality-field-line-following} with the unit vector $\boldsymbol{b}$ to get
\begin{equation}
    {d\ell} =  \frac{dz}{\boldsymbol{b}\cdot \boldsymbol{\nabla}z}.
\label{eqn:gradpar-relation-0}
\end{equation}
Repeating the same process for the coordinate system $(x, y, \hat{z})$ and using~\eqref{eqn:gradpar-relation-0}
\begin{equation}
    \frac{dz}{\boldsymbol{b}\cdot \boldsymbol{\nabla}z} =  \frac{d\hat{z}}{\boldsymbol{b}\cdot \boldsymbol{\nabla}\hat{z}}.
\label{eqn:gradpar-relation-1}
\end{equation}
Integrating~\eqref{eqn:gradpar-relation-1} first for one period in $z$ (or $\hat{z}$) and using the fact that
\begin{equation}
    \boldsymbol{b}\cdot\boldsymbol{\nabla}z \lvert_{z = 0, L} = \boldsymbol{b}\cdot\boldsymbol{\nabla}{\hat{z}} \lvert_{\hat{z} = 0, L},
\end{equation}
we get
\begin{equation}
    C = \boldsymbol{b}\cdot\boldsymbol{\nabla} \hat{z} = L \left(\int_{0}^{L} \frac{dz}{\boldsymbol{b}\cdot \boldsymbol{\nabla}z}\right)^{-1}
\end{equation}
Using $C$, we can now obtain $\hat{z}$ for each value of $z$ by integrating~\eqref{eqn:gradpar-relation-1}
\begin{equation}
    \hat{z} =  C \int_{0}^{L}\frac{dz}{\boldsymbol{b}\cdot \boldsymbol{\nabla}z}. 
\label{eq:gradpar-relation-2}
    \end{equation}
The resulting $\hat{z}$ coordinate is an equal-arc coordinate. Upon obtaining the equal-arc $\hat{z}$, we interpolate all geometric coefficients onto an equispaced grid $\tilde{z}$. This makes $\tilde{z}$ an equispaced, equal-arc coordinate. Note that this procedure can be used for any field-line-following coordinate. 


{\color{black}
\section{Turbulent fluxes}
\label{app:fluxes}
Here we document expressions for turbulent fluxes in the Fourier-Laguerre-Hermite basis.
\subsection{Heat flux}
The normalized radial heat flux for species $s$ is defined as
\begin{align}
\frac{Q_s}{n_s \tau_s} &= \int \frac{dx\, dy\, dz}{\nabla x\times\nabla y\cdot\nabla z}\int d^3{\bf v}\ \gyavgR{\vEM} \cdot\nabla x
\left(\frac{1}{2}v_\parallel^2 + \mu B\right) h_s \notag \\
&= \int \frac{dz}{\nabla x\times\nabla y\cdot\nabla z}\sum_{{\bf k}_\perp} \{i k_y \Phi({\bf k_\perp},z)\}^*\int d^3{\bf v}\left(\frac{1}{2}v_\parallel^2 + \mu B\right) J_{0s} h_s \notag \\
&\quad - v_{ts} \int \frac{dz}{\nabla x\times\nabla y\cdot\nabla z}\sum_{{\bf k}_\perp} \{i k_y A_\parallel({\bf k_\perp},z)\}^* \int d^3{\bf v} \left(\frac{1}{2}v_\parallel^2 + \mu B\right)  J_{0s} v_\parallel h_s \notag \\
&\quad + \frac{\tau_s}{Z_s} \int \frac{dz}{\nabla x\times\nabla y\cdot\nabla z}\sum_{{\bf k}_\perp}   \left\{ik_y\frac{\delta\! B_\parallel({\bf k_\perp},z)}{B}\right\}^* \int d^3{\bf v} \left(\frac{1}{2}v_\parallel^2 + \mu B\right) 2\mu B \frac{J_{1s}}{\alpha_s} h_s \notag \\
&= \int \frac{dz}{\nabla x\times\nabla y\cdot\nabla z}\sum_{{\bf k}_\perp} \{i k_y \Phi({\bf k_\perp},z)\}^*\, \frac{3}{2}\bar{p}_s 
- v_{ts} \int \frac{dz}{\nabla x\times\nabla y\cdot\nabla z}\sum_{{\bf k}_\perp} \{i k_y A_\parallel({\bf k_\perp},z)\}^*\, \bar{q}_{\parallel s}\notag \\
&\quad + \frac{\tau_s}{Z_s} \int \frac{dz}{\nabla x\times\nabla y\cdot\nabla z}\sum_{{\bf k}_\perp}   \left\{ik_y\frac{\delta\! B_\parallel({\bf k_\perp},z)}{B}\right\}^*\, \bar{q}_{\perp s}
\end{align} 
where $\{...\}^*$ denotes complex conjugate, and
\begin{gather}
    \frac{3}{2}\bar{p}_s = \sum_{\ell=0}^{N_\ell} \left\{ 
  \left[\ell \mathcal{J}^s_{\ell-1}+ (2\ell+\frac{3}{2})\mathcal{J}^s_\ell
     + (\ell+1)\mathcal{J}^s_{\ell+1}\right] H^s_{\ell,0} + \frac{1}{\sqrt{2}}\mathcal{J}^s_\ell H^s_{\ell,2} \right\} \\
     \bar{q}_{\parallel s} = \sum_{\ell=0}^{N_\ell} \left\{\left[\ell \mathcal{J}^s_{\ell-1}+ (2\ell+\frac{5}{2})\mathcal{J}^s_\ell
     + (\ell+1)\mathcal{J}^s_{\ell+1}\right] H^s_{\ell,1}  +  \sqrt{\frac{3}{2}} \mathcal{J}^s_\ell H^s_{\ell,3}  \right\} \\
     \bar{q}_\perp = \sum_{\ell=0}^{N_\ell} \left\{ 
  \left[\ell \mathcal{J}^s_{\ell-2} + (3\ell+\frac{3}{2})\mathcal{J}^s_{\ell-1}+ (3\ell+\frac{5}{2})\mathcal{J}^s_\ell 
     + (\ell+1)\mathcal{J}^s_{\ell+1}\right] H^s_{\ell,0} + \frac{1}{\sqrt{2}}(\mathcal{J}^s_\ell + \mathcal{J}^s_{\ell-1}) H^s_{\ell,2} \right\}
\end{gather}
Note that since $\vEM \cdot \nabla \chi = 0$, the convection of the
integral of $h$ at fixed ${\bf r}$ is equal to the convection of the
integral of $g$ at fixed ${\bf r}$. In physical units, the heat flux for species $s$ is $Q_{{\rm phys},
  s} = (n_0 v_t T)_{\rm ref} \rho_*^2 Q_s\equiv Q_{GB,\mathrm{ref}} Q_s$.

\subsection{Particle flux}
Similarly, the normalized radial particle flux for species $s$ is defined as
\begin{align}
\frac{\Gamma_s}{n_s} &= \int \frac{dx\, dy\, dz}{\nabla x\times\nabla y\cdot\nabla z}\int d^3{\bf v}\ \gyavgR{\vEM} \cdot\nabla x
 \,h \notag \\
&= \int \frac{dz}{\nabla x\times\nabla y\cdot\nabla z}\sum_{{\bf k}_\perp} \{i k_y \Phi({\bf k_\perp},z)\}^*\int d^3{\bf v} J_{0s} h_s \notag \\
&\quad - v_{ts} \int \frac{dz}{\nabla x\times\nabla y\cdot\nabla z}\sum_{{\bf k}_\perp} \{i k_y A_\parallel({\bf k_\perp},z)\}^* \int d^3{\bf v} J_{0s} v_\parallel h_s \notag \\
&\quad + \frac{\tau_s}{Z_s} \int \frac{dz}{\nabla x\times\nabla y\cdot\nabla z}\sum_{{\bf k}_\perp}   \left\{ik_y\frac{\delta\! B_\parallel({\bf k_\perp},z)}{B}\right\}^* \int d^3{\bf v}\,  2\mu B \frac{J_{1s}}{\alpha_s} h_s \notag \\
&= \int \frac{dz}{\nabla x\times\nabla y\cdot\nabla z}\sum_{{\bf k}_\perp} \{i k_y \Phi({\bf k_\perp},z)\}^*\, \bar{n}_s \notag \\
&\quad - v_{ts} \int \frac{dz}{\nabla x\times\nabla y\cdot\nabla z}\sum_{{\bf k}_\perp} \{i k_y A_\parallel({\bf k_\perp},z)\}^*\, \bar{u}_{\parallel s}\notag \\
&\quad + \frac{\tau_s}{Z_s} \int \frac{dz}{\nabla x\times\nabla y\cdot\nabla z}\sum_{{\bf k}_\perp}   \left\{ik_y\frac{\delta\! B_\parallel({\bf k_\perp},z)}{B}\right\}^*\, \frac{\bar{u}_{\perp s}}{\sqrt{b_s}}
\end{align} 
with 
\begin{gather}
\bar{n}_s = \sum_{\ell=0}^{N_\ell}  \gm{}{\ell}^s H_{\ell,0}^s, \\ 
\bar{u}_{\parallel s} = \sum_{\ell=0}^{\color{black} N_\ell} \gm{}{\ell}^s H_{\ell,1}^s, \label{eq:ubarHLflux}\\
\frac{\bar{u}_{\perp s}}{\sqrt{b_s}} =  \sum_{\ell=0}^{\color{black} N_\ell}
\left(\mathcal{J}_\ell^s+\mathcal{J}_{\ell-1}^s\right)H^s_{\ell,0}, 
\end{gather}
as defined in \cref{sec:lh}.
In physical units, the particle flux for species $s$ is $\Gamma_{{\rm
    phys}, s} = (n_0 v_t)_{\rm ref} \rho_*^2 \Gamma_s\equiv \Gamma_{GB,\mathrm{ref}}\Gamma_s$.
}

\section{Numerical resolution for benchmark cases} \label{app:resolution}
Here we provide the numerical resolution and associated parameters used in each benchmark case from \cref{sec:benchmarks}. In all cases, normalizing quantities below follow the \GX conventions above (as opposed to the normalization conventions of the various other codes).

\subsection{Linear Cyclone Base Case, Boltzmann electrons (\cref{fig:lin_adiab})}
\GX calculations used: three $2\pi$ segments in an extended ballooning domain each with $N_z=24$ parallel grid points, $N_m=48$ Hermite modes, and $N_\ell=16$ Laguerre modes. 

GS2 calculations used: three $2\pi$ segments in an extended ballooning domain each with $N_z=25$ parallel grid points, 32 energy grid points, and 33 pitch angles (20 in the untrapped region of velocity space and 13 in the trapped region). 

\subsection{Linear Cyclone Base Case, kinetic electrons (\cref{fig:lin_kin} \& \cref{fig:itg_kbm})}
\GX calculations used: three $2\pi$ segments in an extended ballooning domain each with $N_z=24$ parallel grid points, $N_m=128$ Hermite modes, and $N_\ell=16$ Laguerre modes.

GS2 calculations used: three $2\pi$ segments in an extended ballooning domain each with $N_z=25$ parallel grid points, 32 energy grid points, and 33 pitch angles (20 in the untrapped region of velocity space and 13 in the trapped region).

stella calculations used: three $2\pi$ segments in an extended ballooning domain each with $N_z=25$ parallel grid points, 48 $v_\parallel$ (evenly-spaced) grid points with $v_{\parallel,\mathrm{max}}=3\sqrt{2} v_{ti}$, and 16 $\mu$ grid points with $v_{\perp,\mathrm{max}} = 3\sqrt{2} v_{ti}$. 

\subsection{Nonlinear Cyclone Base Case, Boltzmann electrons (\cref{fig:nl_cyclone} and \cref{tab:nl_cyclone})} \label{app:resolution_nl_cyclone}
\GX calculations used: one $2\pi$ ballooning domain segment with $N_z=24$ parallel grid points, $N_x=192$ radial grid points (which corresponds to $N_{k_x}=127$ dealiased Fourier modes), $N_y=64$ binormal grid points (which corresponds to $N_{k_y}=22$ dealiased $k_y\geq0$ Fourier modes, with the $k_y<0$ modes determined by the reality condition). The velocity-space resolution varied by case and is provided in the main text. The perpendicular box dimensions were approximately $190\rho_i\times 190\rho_i$. The hyper-dissipation parameters used were: hyper-viscosity with $D=0.05$ and $n=4$; hyper-collisions with $f_\mathrm{hyp}=1$ and $p=N_m/2$.

GS2 calculations used: one $2\pi$ ballooning domain segment with $N_z=24$ parallel grid points, $N_x=192$ radial grid points ($N_{k_x}=127$), $N_y=64$ binormal grid points ($N_{k_y}=22$), 16 energy grid points, and 33 pitch angles (20 in the untrapped region of velocity space and 13 in the trapped region). The perpendicular box dimensions were approximately $190\rho_i\times 190\rho_i$.

\subsection{Nonlinear Cyclone Base Case, kinetic electrons (\cref{fig:nl_cyclone_kin} and \cref{tab:nl_cyclone_kin})}\label{app:resolution_nl_cyclone_kin}

\GX calculations used: one $2\pi$ ballooning domain segment with $N_z=24$ parallel grid points, $N_x=192$ radial grid points (which corresponds to $N_{k_x}=127$ dealiased Fourier modes), $N_y=64$ binormal grid points (which corresponds to $N_{k_y}=22$ dealiased Fourier modes). The velocity-space resolution varied by case and is provided in the main text. The perpendicular box dimensions were approximately $190\rho_i\times 190\rho_i$. The hyper-dissipation parameters used were: hyper-viscosity with $D=0.05$ and $n=4$; hyper-collisions with $f_\mathrm{hyp}=1$ and $p=N_m/2$.

GS2 calculations used: one $2\pi$ ballooning domain segment with $N_z=32$ parallel grid points, $N_x=192$ radial grid points ($N_{k_x}=127$), $N_y=64$ binormal grid points ($N_{k_y}=22$), 16 energy grid points, and 36 pitch angles (19 in the untrapped region of velocity space and 17 in the trapped region). The perpendicular box dimensions were approximately $190\rho_i\times 190\rho_i$.

\subsection{Linear W7-X case (\cref{fig:lin_w7x})}

\GX calculations used: six $2\pi$ ballooning domain segments (corresponding to six poloidal turns) with $N_z=256$ parallel grid points, $N_m=16$ Hermite modes, and $N_\ell=8$ Laguerre modes.

stella calculations used: 33 field periods with $N_z=256$ parallel grid points, $N_{v_\parallel}=32$, and $N_\mu=8$.

\subsection{Nonlinear W7-X case (\cref{fig:nl_w7x})}

\GX calculations used: one $2\pi$ ballooning domain segment (corresponding to a single poloidal turn) with $N_z=128$ parallel grid points, $N_x=64$ radial grid points (which corresponds to $N_{k_x}=43$ dealiased Fourier modes), $N_y=192$ binormal grid points (which corresponds to $N_{k_y}=64$ dealiased Fourier modes). The velocity-space resolution varied by case and is provided in the main text. The perpendicular box dimensions were approximately $132\rho_i\times 89\rho_i$. The hyper-dissipation parameters used were: hyper-viscosity with $D=0.05$ and $n=4$; hyper-collisions with $f_\mathrm{hyp}=1$ and $p=N_m/2$.

For stella and GENE calculation details, refer to Tables 2 and 3 (test 5) of \citet{Gonzalez-Jerez2022}.

\bibliographystyle{jpp}
\bibliography{library}

\end{document}